\documentclass{article}
\usepackage{amsmath}
\usepackage{amssymb}

\input{tcilatex}
\begin{document}

\begin{center}
462PolsWithMultiplZerosAndSolvDynSyst181216Rev\bigskip

{\Huge Polynomials with Multiple Zeros and Solvable Dynamical Systems
including Models in the Plane with Polynomial Interactions}

\bigskip

\textbf{Francesco Calogero}$^{a,b,1}${\LARGE \ {\large and} }\textbf{Farrin
Payandeh}$^{a,c,2}$

$^{a}$ Physics Department, University of Rome "La Sapienza", Rome, Italy

$^{b}$ INFN, Sezione di Roma 1

$^{c}$ Department of Physics, Payame Noor University (PNU), PO BOX
19395-3697 Tehran, Iran

$^{1}$ francesco.calogero@roma1.infn.it, francesco.calogero@uniroma1.it

$^{2}$ f\_payandeh@pnu.ac.ir, farrinpayandeh@yahoo.com

\bigskip

\textit{Abstract}
\end{center}

The interplay among the time-evolution of the \textit{coefficients} $%
y_{m}\left( t\right) $ and the \textit{zeros} $x_{n}\left( t\right) $ of a
generic time-dependent (monic) polynomial provides a convenient tool to
identify certain classes of \textit{solvable} dynamical systems. Recently
this tool has been extended to the case of \textit{nongeneric} polynomials
characterized by the presence, for \textit{all} time, of a \textit{single}
\textit{double} zero; and subsequently significant progress has been made to
extend this finding to the case of polynomials featuring a \textit{single}
zero of \textit{arbitrary} multiplicity. In this paper we introduce an
approach suitable to deal with the most general case, i. e. that of a
\textit{nongeneric} time-dependent polynomial with an \textit{arbitrary}
number of zeros each of which features, for all time, an \textit{arbitrary}
(time-independent) multiplicity. We then focus on the special case of a
polynomial of degree $4$ featuring only $2$ different zeros and, by using a
recently introduced additional twist of this approach, we thereby identify
many new classes of \textit{solvable} dynamical systems of the following
type:%
\begin{equation*}
\dot{x}_{n}=P^{\left( n\right) }\left( x_{1},x_{2}\right) ~,~~~n=1,2~,
\end{equation*}%
with $P^{\left( n\right) }\left( x_{1},x_{2}\right) $ two \textit{polynomials%
} in the two variables $x_{1}\left( t\right) $ and $x_{2}\left( t\right) $.

\bigskip

\section{Introduction}

\textbf{Notation 1.1}. Hereafter $t$ generally denotes \textit{time} (the
\textit{real} independent variable); (partial) derivatives with respect to
time are denoted by a superimposed dot, or in some case by appending as a
subscript the independent variable $t$ preceded by a comma; all dependent
variables such as $x$, $y$, $z$ (often equipped with subscripts) are
generally assumed to be \textit{complex} numbers, unless otherwise indicated
(it shall generally be clear from the context which of these and other
quantities depend on the time $t$, as occasionally---but not always---%
\textit{explicitly} indicated); parameters such as $a$, $b,$ $c,$ $\alpha ,$
$\beta ,$ $\gamma ,$ $A$, etc. (often equipped with subscripts) are
generally time-independent \textit{complex} numbers; and indices such as $n$%
, $m$, $j$, $\ell $ are generally \textit{positive integers}, with ranges
explicitly indicated or clear from the context. $\blacksquare $

Some time ago the idea has been exploited to identify dynamical systems
which can be solved by using as a \textit{tool} the relations between the
time evolutions of the \textit{coefficients} and the \textit{zeros} of a
\textit{generic} \textit{time-dependent} polynomial \cite{C1978}. The basic
idea of this approach is to relate the time-evolution of the $N$ \textit{%
zeros} $x_{n}\left( t\right) $ of a \textit{generic} time-dependent
polynomial $p_{N}\left( z;t\right) $ of degree $N$ in its argument $z,$%
\begin{equation}
p_{N}\left( z;t\right) =z^{N}+\sum_{m=1}^{N}\left[ y_{m}\left( t\right)
z^{N-m}\right] =\prod_{n=1}^{N}\left[ z-x_{n}\left( t\right) \right] ~,
\label{1pNzt}
\end{equation}%
to the time-evolution of its $N$ \textit{coefficients} $y_{m}\left( t\right)
$. Indeed, if the time evolution of the $N$ \textit{coefficients} $%
y_{m}\left( t\right) $ is determined by a system of Ordinary Differential
Equations (ODEs) which is itself \textit{solvable}, then the corresponding
time-evolution of the $N$ \textit{zeros} $x_{n}\left( t\right) $ is also
\textit{solvable}, via the following $3$ steps: (i) given the initial values
$x_{n}\left( 0\right) ,$ the corresponding initial values $y_{m}\left(
0\right) $ can be obtained from the \textit{explicit} formulas expressing
the \textit{coefficients} $y_{m}$ of a polynomial in terms of its \textit{%
zeros} $x_{n}$ reading (for all time, hence in particular at $t=0$)%
\begin{equation}
y_{m}\left( t\right) =\left( -1\right) ^{m}\sum_{1\leq
n_{1}<n_{2}<...<n_{m}\leq N}^{N}\left\{ \prod_{\ell =1}^{M}\left[ x_{n_{\ell
}}\left( t\right) \right] \right\} ~,~~~m=1,2,...,N~;  \label{1ymxn}
\end{equation}%
(ii) from the $N$ values $y_{m}\left( 0\right) $ thereby obtained, the $N$
values $y_{m}\left( t\right) $ are then evaluated via the---assumedly
\textit{solvable}---system of ODEs satisfied by the $N$ coefficients $%
y_{m}\left( t\right) $; (iii) the $N$ values $x_{n}\left( t\right) $---i.
e., the $N$ solutions of the dynamical system satisfied by the $N$ variables
$x_{n}\left( t\right) $---are then determined as the $N$ zeros of the
polynomial, see (\ref{1pNzt}), itself known at time $t$ in terms of its $N$
coefficients $y_{m}\left( t\right) $ (the computation of the \textit{zeros}
of a known polynomial being an \textit{algebraic} operation; of course
generally \textit{explicitly} performable only for polynomials of degree $%
N\leq 4$).

\textbf{Remark 1-1}. In this paper the term "solvable" generally
characterizes systems of ODEs the initial-values of which are "solvable by
algebraic operations"---possibly including quadratures yielding implicit
solutions, generally also requiring the evaluation of parameters via \textit{%
algebraic} operations. And let us emphasize that, because of the \textit{%
algebraic} but \textit{nonlinear} character of the relations between the
\textit{zeros} and the \textit{coefficients} of a polynomial, it is clear
that, even to relatively \textit{trivial} evolutions of the $N$ \textit{%
coefficients} $y_{m}\left( t\right) $ of a time-dependent polynomial, there
correspond \textit{much less trivial} evolutions of its $N$ zeros $%
x_{n}\left( t\right) $. On the other hand the fact that a time evolution is
\textit{algebraically solvable} has important implications, generally
excluding that it can be "chaotic", indeed in some cases allowing to infer
important qualitative features of the time evolution, such as the property
to be \textit{isochronous} or \textit{asymptotically isochronous} (see for
instance \cite{C2008} \cite{GS2005} \cite{CG2008}). $\blacksquare $

The viability of this technique to identify \textit{solvable} dynamical
systems depends of course on the availability of an \textit{explicit} method
to relate the time-evolution of the $N$ zeros of a \textit{polynomial} to
the corresponding time-evolution of its $N$ \textit{coefficients}. Such a
method was indeed provided in \cite{C1978}, opening the way to the
identification of a vast class of \textit{algebraically solvable} dynamical
systems (see also \cite{C2001} and references therein); but that approach
was essentially restricted to the consideration of \textit{linear} time
evolutions of the coefficients $y_{m}\left( t\right) $.

A development allowing to lift this quite strong restriction emerged
relatively recently \cite{C2016}, by noticing the validity of the \textit{%
identity}%
\begin{equation}
\dot{x}_{n}=-\left[ \prod_{\ell =1,~\ell \neq n}^{N}\left( x_{n}-x_{\ell
}\right) \right] ^{-1}\sum_{m=1}^{N}\left[ \dot{y}_{m}\left( x_{n}\right)
^{N-m}\right]  \label{1xndot}
\end{equation}%
which provides a convenient \textit{explicit} relationship among the time
evolutions of the $N$ \textit{zeros} $x_{n}\left( t\right) $ and the $N$
\textit{coefficients} $y_{m}\left( t\right) $ of the generic polynomial (\ref%
{1pNzt}). This allowed a major enlargement of the class of \textit{%
algebraically solvable} dynamical systems identifiable via this approach:
for many examples see \cite{C2018} and references therein.

\textbf{Remark 1-2}. Analogous identities to (\ref{1xndot}) have been
identified for higher time-derivatives \cite{BC2016a} \cite{BC2016b} \cite%
{C2018}; but in this paper we restrict our treatment to dynamical systems
characterized by \textit{first-order} ODEs, postponing the treatment of
dynamical systems characterized by \textit{higher-order} ODEs (see \textbf{%
Section 6}). $\blacksquare $

A new twist of this approach was then provided by its extension to \textit{%
nongeneric} polynomials featuring---for \textit{all} time---\textit{multiple}
zeros. The first step in this direction focussed on time-dependent
polynomials featuring for \textit{all} time a \textit{single double zero}
\cite{BC2018}; and subsequently significant progress has been made to treat
the case of polynomials featuring a \textit{single zero} of \textit{%
arbitrary multiplicity} \cite{B2018}. In \textbf{Section 2} of the present
paper a convenient method is provided which is suitable to treat the most
general case of polynomials featuring an \textit{arbitrary} number of
\textit{zeros} each of which features an \textit{arbitrary multiplicity}.
While all these developments might appear to mimic scholastic exercises
analogous to the discussion among medieval scholars of how many angels might
dance simultaneously on the point of\ a needle, they do indeed provide
\textit{new tools} to identify \textit{new} dynamical systems featuring
interesting time evolutions (including systems displaying remarkable
behaviors such as \textit{isochrony} or \textit{asymptotic isochrony}: see
for instance \cite{BC2018} \cite{B2018}); dynamical systems which---besides
their intrinsic mathematical interest---are quite likely to play significant
roles in applicative contexts.

Such developments shall be reported in future publications. In the present
paper we focus on another twist of this approach to identify new \textit{%
solvable} dynamical systems which was introduced quite recently \cite{CP2018}%
. It is again based on the relations among the time-evolution of the \textit{%
coefficients} and the \textit{zeros} of time-dependent polynomials \cite%
{C2016} \cite{C2018} with \textit{multiple roots} (see \cite{BC2018}, \cite%
{B2018} and above); but (as in \cite{CP2018}) by restricting attention to
such polynomials featuring \textit{only }$2$ \textit{zeros}. Again, this
might seem such a strong limitation to justify the doubt that the results
thereby obtained be of much interest. But the effect of this restriction is
to open the possibility to identify \textit{algebraically solvable }%
dynamical models characterized by the following system of $2$ ODEs,%
\begin{equation}
\dot{x}_{n}=P^{\left( n\right) }\left( x_{1},x_{2}\right) ~,~~~n=1,2~,
\label{1xndotPol}
\end{equation}%
with $P^{\left( n\right) }\left( x_{1},x_{2}\right) $ two \textit{polynomials%
} in the two dependent variables $x_{1}\left( t\right) $ and $x_{2}\left(
t\right) $; hence systems of considerable interest, both from a theoretical
and an applicative point of view (see \cite{CP2018} and references quoted
there). This development is detailed in the following \textbf{Section 3} by
treating a specific example. In \textbf{Section 4} we report---without
detailing their derivation, which is rather obvious on the basis of the
treatment provided in \textbf{Section 3}---many other such \textit{solvable}
models (see (\ref{1xndotPol}); but in some cases the right-hand side of
these equations are not quite polynomial); and a simple technique allowing
additional extensions of these models---making them potentially more useful
in applicative contexts---is outlined in \textbf{Section 5}, by detailing
its applicability in a particularly interesting case. Hence researchers
primarily interested in \textit{applications} of such systems of ODEs might
wish to take first of all a quick look at these $2$ sections.

Finally, \textbf{Section 6} outlines future developments of this research
line; and some material useful for the treatment provided in the body of
this paper is reported in $2$ Appendices.

\bigskip

\section{Properties of nongeneric time-dependent polynomials featuring $N$
zeros, each of \textit{arbitrary} multiplicity}

In this \textbf{Section} 2 we focus on time-dependent (monic) polynomials
featuring for all time $N$ different zeros $x_{n}\left( t\right) $, each of
which with the \textit{arbitrarily} assigned (of course time-independent)
multiplicity $\mu _{n}$. They are of course defined as follows:
\begin{subequations}
\label{2PMzt}
\begin{equation}
P_{M}\left( z;t\right) =z^{M}+\sum_{m=1}^{M}\left[ y_{m}\left( t\right)
z^{M-m}\right] =\prod_{n=1}^{N}\left\{ \left[ z-x_{n}\left( t\right) \right]
^{\mu _{n}}\right\} ~.  \label{2PMzta}
\end{equation}%
Here the $N$ \textit{positive integers} $\mu _{n}$ are \textit{a priori}
arbitrary. It is obvious that this formula implies that the degree of this
polynomial $P_{M}\left( z;t\right) $ is%
\begin{equation}
M=\sum_{n=1}^{N}\left( \mu _{n}\right) ~.  \label{2M}
\end{equation}

It is plain that there exist \textit{explicit} formulas---generalizing (\ref%
{1ymxn})---expressing the $M$ coefficients $y_{m}\left( t\right) $ in terms
of the $N$ zeros $x_{n}\left( t\right) $; for instance clearly
\end{subequations}
\begin{equation}
y_{1}\left( t\right) =-\sum_{n=1}^{N}\left[ \mu _{n}x_{n}\left( t\right) %
\right] ~,~~~y_{M}\left( t\right) =\left( -1\right)
^{M}\prod_{n=1}^{N}\left\{ \left[ x_{n}\left( t\right) \right] ^{\mu
_{n}}\right\} ~,
\end{equation}%
and see other examples below.

It is also plain that, while $N$ \textit{zeros} $x_{n}$ and their
multiplicities $\mu _{n}$ can be \textit{arbitrarily} assigned in order to
define the polynomial (\ref{2PMzt}), this is \textit{not} the case for the $%
M $ \textit{coefficients} $y_{m}$: generally---for any given assignment of
the $N$ multiplicities $\mu _{n}$---only $N$ of them can be \textit{%
arbitrarily} assigned, thereby determining (via algebraic operations) the $N$
zeros $x_{n} $ and the remaining $M-N$ other coefficients $y_{m}.$

\textbf{Remark 2-1}. The \textit{generic} polynomial (\ref{1pNzt}), of
degree $N$ and featuring $N$ different zeros $x_{n}$ and $N$ coefficients $%
y_{m}$, generally implies that the set of its $N$ coefficients $y_{m}$ is an
$N$-vector $\vec{y}=\left( y_{1},...,y_{N}\right) ,$ while the set of its $N$
zeros $x_{n}$ is instead an \textit{unordered} set of $N$ numbers $x_{n}$.
This however is not quite true in the case of a \textit{time-dependent}
generic polynomial which features---as those generally considered in this
paper---a \textit{continuous} time-dependence of its \textit{coefficients}
and \textit{zeros}; then the set of its $N$ zeros $x_{n}\left( 0\right) $ at
the initial time $t=0$ should be generally considered an \textit{unordered}
set, but for all subsequent time, $t>0$, the set of its $N$ zeros $%
x_{n}\left( t\right) $ is an \textit{ordered} set, the assignment of the
index $n$ to $x_{n}\left( t\right) $ being no more arbitrary but rather
\textit{determined by continuity} in $t$ (at least provided during the time
evolution no collision of two or more \textit{different zeros} occur, in
which case the identities of \textit{these zeros} get to some extent lost
because their identities may be exchanged, becoming \textit{undetermined}).

The situation is quite different in the case of a \textit{nongeneric}
polynomial such as (\ref{2PMzt}): then zeros having \textit{different}
multiplicities are intrinsically different, for instance if all the
multiplicities $\mu _{n}$ are different among themselves, $\mu _{n}\neq \mu
_{\ell }$ if $n\neq \ell $, then clearly the set of the $N$ zeros $x_{n}$ is
an \textit{ordered} set (hence an $N$-vector).

We trust the reader to understand these rather obvious facts and therefore
hereafter we refrain from any additional discussion of these issues. $%
\blacksquare $

Our task now is to identify---for the special class of \textit{nongeneric}
polynomials (\ref{2PMzt})--- equivalent relations to the identities (\ref%
{1xndot}), to be then used in order to identify \textit{new solvable}
dynamical systems.

The first step is to time-differentiate once the formula (\ref{2PMzt}),
getting the relations%
\begin{eqnarray}
&&P_{M,t}\left( z;t\right) =\sum_{m=1}^{M}\left[ \dot{y}_{m}\left( t\right)
z^{M-m}\right]  \notag \\
&=&-\sum_{n=1}^{N}\mu _{n}\dot{x}_{n}\left( t\right) \left[ z-x_{n}\left(
t\right) \right] ^{\mu _{n}-1}\prod_{\ell =1,~\ell \neq n}^{N}\left\{ \left[
z-x_{\ell }\left( t\right) \right] ^{\mu _{\ell }}\right\} ~.  \label{2Pmtzt}
\end{eqnarray}

\textbf{Remark 2-2}. Hereafter, in order to avoid clattering our
presentation with unessential details, we occasionally make the convenient
assumption that all the numbers $\mu _{n}$ be \textit{different} among
themselves; the diligent reader shall have no difficulty to understand how
the treatment can be extended to include cases in which this simplifying
assumption does not hold---indeed in the specific examples discussed below
we will include in our treatment also cases in which this simplification is
not valid, taking appropriate care of such cases. And we also
assume---without loss of generality---that the numbers $\mu _{n}$ are
ordered in decreasing order, $\mu _{n}\geq \mu _{n+1}$. $\blacksquare $

Our next step is to $z$-differentiate $\mu $ times the above formulas,
firstly with $\mu =0,1,2,...,\mu _{n}-2$ and secondly with $\mu =\mu _{n}-1$%
; and then set $z=x_{n}$ (for each value of $n=1,2,...,N$). There clearly
thereby obtain the following formulas:
\begin{subequations}
\label{2Eqydot}
\begin{eqnarray}
&&\sum_{m=1}^{M-\mu }\left\{ \dot{y}_{m}\left( t\right) \left[ \frac{\left(
M-m\right) !}{\left( M-m-\mu \right) !}\right] \left[ x_{n}\left( t\right) %
\right] ^{M-m-\mu }\right\} =0~,  \notag \\
\mu  &=&0,1,...,\mu _{n}-2~,~~~n=1,2,...,N~;  \label{2Eqydot0}
\end{eqnarray}

\begin{eqnarray}
&&\sum_{m=1}^{M-\mu _{n}+1}\left\{ \dot{y}_{m}\left( t\right) \left[ \frac{%
\left( M-m\right) !}{\left( M-m-\mu _{n}+1\right) !}\right] \left[
x_{n}\left( t\right) \right] ^{M-m-\mu _{n}+1}\right\}  \notag \\
&=&-\left( \mu _{n}!\right) \dot{x}_{n}\left( t\right) \prod_{\ell =1,~\ell
\neq n}^{N}\left\{ \left[ x_{n}\left( t\right) -x_{\ell }\left( t\right) %
\right] ^{\mu _{\ell }}\right\} ~,  \notag \\
n &=&1,2,...,N~.  \label{2Eqxndot}
\end{eqnarray}

The second set, (\ref{2Eqxndot}), yields the following $N$ expressions of
the time-derivatives of the $N$ zeros $x_{n}\left( t\right) $ in terms of
the time-derivatives of the $M$ coefficients $y_{m}\left( t\right) $:
\end{subequations}
\begin{eqnarray}
&&\dot{x}_{n}\left( t\right) =-\left\{ \mu _{n}!\prod_{\ell =1,~\ell \neq
n}^{N}\left[ x_{n}\left( t\right) -x_{\ell }\left( t\right) \right] ^{\mu
_{\ell }}\right\} ^{-1}\cdot  \notag \\
&&\cdot \sum_{m=1}^{M-\mu _{n}+1}\left\{ \dot{y}_{m}\left( t\right) \left[
\frac{\left( M-m\right) !}{\left( M-m-\mu _{n}+1\right) !}\right] \left[
x_{n}\left( t\right) \right] ^{M-m-\mu _{n}+1}\right\} ~,  \notag \\
&&n=1,2,...,N~.  \label{2xxndot}
\end{eqnarray}

The first set, (\ref{2Eqydot0}), consists of \textit{linear} relations among
the $M$ time-derivatives $\dot{y}_{m}\left( t\right) $ of the $M$
coefficients $y_{m}\left( t\right) $: and it is easily seen, via (\ref{2M}),
that there are altogether%
\begin{equation}
\sum_{n=1}^{N}\left( \mu _{m}-1\right) =M-N
\end{equation}%
such relations. So one can select $N$ quantities $\dot{y}_{m}\left( t\right)
$---let us hereafter call them $\dot{y}_{\tilde{m}}\left( t\right) $---and
compute, from the $M-N$ linear equations (\ref{2Eqydot0}), all the other $M-N
$ quantities $\dot{y}_{m}\left( t\right) $ with $m\neq \tilde{m}$ as \textit{%
linear} expressions in terms of these selected $N$ quantities $\dot{y}_{%
\tilde{m}}\left( t\right) $. The goal of expressing the $N$ time-derivatives
$\dot{x}_{n}$ as \textit{linear} equations---somehow analogous to the
identities (\ref{1xndot})---in terms of the $N$ time-derivatives $\dot{y}_{%
\tilde{m}}\left( t\right) $ of $N$, \textit{arbitrarily} selected,
coefficients $y_{\tilde{m}}\left( t\right) $ is thereby finally achieved.
Indeed the task of expressing the $M-N$ quantities $\dot{y}_{m}\left(
t\right) $ with $m\neq \tilde{m}$ in terms of the $N$ quantities $\dot{y}_{%
\tilde{m}}\left( t\right) $---and of course the $N$ zeros $x_{n}\left(
t\right) $---can in principle be implemented \textit{explicitly} as it
amounts to solving the $M-N$ \textit{linear} equations (\ref{2Eqydot}) for
the $M-N$ unknowns $\dot{y}_{m}\left( t\right) $, with the $N$ quantities $%
\dot{y}_{\tilde{m}}\left( t\right) $ playing there the role of known
quantities; clearly implying that the resulting expressions of the $M-N$
quantities $\dot{y}_{m}\left( t\right) $ are \textit{linear} functions of
the $N$ quantities $\dot{y}_{\tilde{m}}\left( t\right) $. And the insertion
of these \textit{linear} expressions of the $M-N$ quantities $\dot{y}%
_{m}\left( t\right) $ (with $m\neq \tilde{m}$) in terms of the $N$
quantities $\dot{y}_{\tilde{m}}\left( t\right) $ in the $N$ formulas (\ref%
{2xxndot}) fulfils our goal.

The actual implementation of this development must of course be performed on
a case-by-case basis, see below. In the special case with \textit{only one}
multiple zero---and if moreover the indices $\tilde{m}$ are assigned their
\textit{first} $N$ values, i. e. $\tilde{m}=1,2,...,N$---these results shall
reproduce the results of the path-breaking paper \cite{B2018}, which was
confined to the treatment of this special case.

The outcomes of these developments are detailed, in the special case with $%
N=2$ and $M=4$, in the following \textbf{Section 3}; for the motivation of
this drastic restriction see below.

\bigskip

\section{The $N=2,$ $M=4$ case}

In the special case with $N=2$ the formula (\ref{2xxndot}) simplifies,
reading (see (\ref{2M}))
\begin{subequations}
\label{22xndot}
\begin{equation}
\dot{x}_{1}=-\left[ \mu _{1}!\left( x_{1}-x_{2}\right) ^{\mu _{2}}\right]
^{-1}\sum_{m=1}^{1+\mu _{2}}\left\{ \dot{y}_{m}\left[ \frac{\left( \mu
_{1}+\mu _{2}-m\right) !}{\left( \mu _{2}-m+1\right) !}\right] \left(
x_{1}\right) ^{\mu _{2}-m+1}\right\} ~,  \label{22x1dot}
\end{equation}%
\begin{equation}
\dot{x}_{2}=-\left[ \mu _{2}!\left( x_{2}-x_{1}\right) ^{\mu _{1}}\right]
^{-1}\sum_{m=1}^{1+\mu _{1}}\left\{ \dot{y}_{m}\left[ \frac{\left( \mu
_{1}+\mu _{2}-m\right) !}{\left( \mu _{1}-m+1\right) !}\right] \left(
x_{2}\right) ^{\mu _{1}-m+1}\right\} ~.  \label{22x2dot}
\end{equation}

Let us moreover restrict attention to the case with $M=4,$ as the case with $%
M=3$ (implying $\mu _{1}=2,$ $\mu _{2}=1$: see \textbf{Remark 2-2}) has been
already discussed in \cite{BC2018} and \cite{CP2018} and the case with $M=4$
is sufficiently rich (see below) to deserve a full paper.

In the case with $M=4$ there are $2$ possible assignments of the $2$
parameters $\mu _{n}$: (i) $\mu _{1}=3,$ $\mu _{2}=1;$ (ii) $\mu _{1}=\mu
_{2}=2$ (see \textbf{Remark 2-2}, and note that we now include also the case
with $\mu _{1}=\mu _{2}$).

\bigskip

\subsection{Case (i): $\protect\mu _{1}=3,$ $\protect\mu _{2}=1$}

In this case (\ref{2PMzta}) clearly implies the following expressions of the
$4$ \textit{coefficient}s $y_{m}\left( t\right) $ in terms of the $2$
\textit{zeros} $x_{n}\left( t\right) $:
\end{subequations}
\begin{eqnarray}
y_{1} &=&-\left( 3x_{1}+x_{2}\right) ~,~~~y_{2}=3x_{1}\left(
x_{1}+x_{2}\right) ~,  \notag \\
y_{3} &=&-\left( x_{1}\right) ^{2}\left( x_{1}+3x_{2}\right)
~,~~~y_{4}=\left( x_{1}\right) ^{3}x_{2}~.  \label{M4N2mu31ym}
\end{eqnarray}

\textbf{Remark 3.1-1}. Note that these formulas imply that $x_{1}$ and $x_{2}
$ can be computed (in fact, \textit{explicitly}!) from $y_{m_{1}}$ and $%
y_{m_{2}}$---with $m_{1}=1,2,3$ and $m_{1}<m_{2}=2,3,4$) by solving an
\textit{algebraic} equation of degree $m_{2}$. $\blacksquare $

The corresponding equations (\ref{2Eqydot0}) read
\begin{subequations}
\label{22ymdot}
\begin{equation}
\dot{y}_{1}\left( x_{1}\right) ^{3}+\dot{y}_{2}\left( x_{1}\right) ^{2}+\dot{%
y}_{3}x_{1}+\dot{y}_{4}=0~,  \label{22ymdota}
\end{equation}%
\begin{equation}
3\dot{y}_{1}\left( x_{1}\right) ^{2}+2\dot{y}_{2}x_{1}+\dot{y}_{3}=0
\label{22ymdotb}
\end{equation}%
(note that in this case only the formulas with $n=1$ are present).

And the formulas (\ref{22xndot}) read
\end{subequations}
\begin{equation}
\dot{x}_{1}=-\frac{3x_{1}\dot{y}_{1}+\dot{y}_{2}}{3\left( x_{1}-x_{2}\right)
}~,~~\ \dot{x}_{2}=\frac{\left( x_{2}\right) ^{3}\dot{y}_{1}+\left(
x_{2}\right) ^{2}\dot{y}_{2}+x_{2}\dot{y}_{3}+\dot{y}_{4}}{\left(
x_{1}-x_{2}\right) ^{3}}~.  \label{M4N2mu31x12dot}
\end{equation}

There are now $6$ possible different assignments for the indices $\tilde{m}$%
:
\begin{subequations}
\label{M4N2mu13mtilde}
\begin{equation}
\tilde{m}=1,2;~\tilde{m}=1,3;~\tilde{m}=1,4;~\tilde{m}=2,3;~\tilde{m}=2,4;~%
\tilde{m}=3,4~,
\end{equation}%
to which there correspond the $6$ complementary assignments%
\begin{equation}
m=3,4;~m=2,4;~m=2,3;~m=1,4;~m=1,3;~m=1,2~.
\end{equation}

Let us list below the 6 corresponding versions of the ODEs (\ref%
{M4N2mu31x12dot}):
\end{subequations}
\begin{subequations}
\label{M4N2x12dot}
\begin{equation}
\dot{x}_{1}=-\frac{3x_{1}\dot{y}_{1}+\dot{y}_{2}}{3\left( x_{1}-x_{2}\right)
}~,~~\ \dot{x}_{2}=\frac{\left( 2x_{1}+x_{2}\right) \dot{y}_{1}+\dot{y}_{2}}{%
\left( x_{1}-x_{2}\right) }~,  \label{M4N2x12dota}
\end{equation}%
\begin{equation}
\dot{x}_{1}=-\frac{3\left( x_{1}\right) ^{2}\dot{y}_{1}-\dot{y}_{3}}{%
6x_{1}\left( x_{1}-x_{2}\right) }~,~~\ \dot{x}_{2}=\frac{x_{1}\left(
x_{1}+2x_{2}\right) \dot{y}_{1}-\dot{y}_{3}}{2x_{1}\left( x_{1}-x_{2}\right)
}~,
\end{equation}%
\begin{equation}
\dot{x}_{1}=-\frac{\left( x_{1}\right) ^{3}\dot{y}_{1}+\dot{y}_{4}}{3\left(
x_{1}\right) ^{2}\left( x_{1}-x_{2}\right) }~,~~\ \dot{x}_{2}=\frac{\left(
x_{1}\right) ^{2}x_{2}\dot{y}_{1}+\dot{y}_{4}}{\left( x_{1}\right)
^{2}\left( x_{1}-x_{2}\right) }~,
\end{equation}%
\begin{equation}
\dot{x}_{1}=\frac{x_{1}\dot{y}_{2}+\dot{y}_{3}}{3x_{1}\left(
x_{1}-x_{2}\right) }~,~~\ \dot{x}_{2}=-\frac{x_{1}\left( x_{1}+2x_{2}\right)
\dot{y}_{2}+\left( 2x_{1}+x_{2}\right) \dot{y}_{3}}{3\left( x_{1}\right)
^{2}\left( x_{1}-x_{2}\right) }~,
\end{equation}%
\begin{equation}
\dot{x}_{1}=\frac{\left( x_{1}\right) ^{2}\dot{y}_{2}-3\dot{y}_{4}}{6\left(
x_{1}\right) ^{2}\left( x_{1}-x_{2}\right) }~,~~\ \dot{x}_{2}=-\frac{\left(
x_{1}\right) ^{2}x_{2}\dot{y}_{2}-\left( 2x_{1}+x_{2}\right) \dot{y}_{4}}{%
2\left( x_{1}\right) ^{3}\left( x_{1}-x_{2}\right) }~,
\end{equation}%
\begin{equation}
\dot{x}_{1}=-\frac{x_{1}\dot{y}_{3}+3\dot{y}_{4}}{3\left( x_{1}\right)
^{2}\left( x_{1}-x_{2}\right) }~,~~\ \dot{x}_{2}=\frac{x_{1}x_{2}\dot{y}%
_{3}+\left( x_{1}+2x_{2}\right) \dot{y}_{4}}{\left( x_{1}\right) ^{3}\left(
x_{1}-x_{2}\right) }~.
\end{equation}

Next, let us focus to begin with---in order to explain our approach---on the
first, (\ref{M4N2x12dota}), of the $6$ formulas (\ref{M4N2x12dot}). Assume
moreover that the $2$ quantities $y_{1}\left( t\right) $ and $y_{2}\left(
t\right) $ evolve according to the following \textit{solvable} system of
ODEs:
\end{subequations}
\begin{equation}
\dot{y}_{1}=f_{1}\left( y_{1},y_{2}\right) ~,~~~\dot{y}_{2}=f_{2}\left(
y_{1},y_{2}\right) ~.  \label{311y1y2dot}
\end{equation}%
It is then clear---via the identities (\ref{M4N2mu31ym})---that we can
conclude that the dynamical system
\begin{subequations}
\label{311x1x2dot}
\begin{eqnarray}
\dot{x}_{1} &=&-\left[ 3\left( x_{1}-x_{2}\right) \right] ^{-1}\left[
3x_{1}f_{1}\left( -\left( 3x_{1}+x_{2}\right) ,3x_{1}\left(
x_{1}+x_{2}\right) \right) \right.   \notag \\
&&\left. +f_{2}\left( -\left( 3x_{1}+x_{2}\right) ,3x_{1}\left(
x_{1}+x_{2}\right) \right) \right] ~,  \label{311x1dot}
\end{eqnarray}%
\begin{eqnarray}
\dot{x}_{2} &=&\left( x_{1}-x_{2}\right) ^{-1}\left[ \left(
2x_{1}+x_{2}\right) f_{1}\left( -\left( 3x_{1}+x_{2}\right) ,3x_{1}\left(
x_{1}+x_{2}\right) \right) \right.   \notag \\
&&\left. +f_{2}\left( -\left( 3x_{1}+x_{2}\right) ,3x_{1}\left(
x_{1}+x_{2}\right) \right) \right] ~,  \label{311x2dot}
\end{eqnarray}%
is as well \textit{solvable}.

While this is in itself an interesting result---to become more significant
for explicit assignments of the $2$ functions $f_{1}\left(
y_{1},y_{2}\right) $ and $f_{2}\left( y_{1},y_{2}\right) $ (see below)---an
additional interesting development emerges if---following the approach of
\cite{CP2018}---we now assume the $2$ functions $f_{1}\left(
y_{1},y_{2}\right) $ and $f_{2}\left( y_{1},y_{2}\right) $ to be \textit{%
both polynomial} in their $2$ arguments and moreover such that
\end{subequations}
\begin{equation}
3xf_{1}\left( -4x,6x^{2}\right) +f_{2}\left( -4x,6x^{2}\right) =0\ ;
\label{31ConPol}
\end{equation}%
a restriction that is clearly \textit{sufficient} to guarantee that the
right-hand sides of the equations of motion (\ref{311x1x2dot}) become
\textit{polynomials} in the $2$ dependent variables $x_{1}\left( t\right) $
and $x_{2}\left( t\right) $ (since the numerators in the right-hand sides of
the $2$ ODEs (\ref{311x1x2dot}) are then both \textit{polynomials} in the
variables $x_{1}$ and $x_{2}$ which vanish when $x_{1}=x_{2}=x$ and which
therefore contain the factor $x_{1}-x_{2}$).

An representative example of such functions is
\begin{subequations}
\label{M4N2f12}
\begin{equation}
f_{1}\left( y_{1},y_{2}\right) =\alpha _{0}+\alpha _{1}y_{2}~,~~~f_{2}\left(
y_{1},y_{2}\right) =\beta _{0}y_{1}+\beta _{1}\left( y_{1}\right) ^{3}~,
\label{M4N2f12a}
\end{equation}%
with (see (\ref{31ConPol}))%
\begin{equation}
\alpha _{0}=\frac{4\beta _{0}}{3}~,~~~\alpha _{1}=\frac{32\beta _{1}}{9}~;
\label{M4N2falphabeta}
\end{equation}%
note that the corresponding equations of motion (\ref{311y1y2dot}) are then
indeed \textit{solvable,} see---up to trivial rescalings of some
parameters---the solution in terms of Jacobian elliptic functions in \textbf{%
Example 1} in \cite{CP2018}, and, below, in \textbf{Subsection} \textbf{Case
A.3.1 }of \textbf{Appendix A}.

The conclusion is then that the dynamical system
\end{subequations}
\begin{subequations}
\label{311x1x2dot}
\begin{equation}
\dot{x}_{1}=a+b\left[ 5\left( x_{1}\right) ^{2}+10x_{1}x_{2}+\left(
x_{2}\right) ^{2}\right] ~,
\end{equation}%
\begin{equation}
\dot{x}_{2}=a+b\left[ 17\left( x_{1}\right) ^{2}+2x_{1}x_{2}-3\left(
x_{2}\right) ^{2}\right] ~,
\end{equation}%
with $a=-\beta _{0}/3$ and $b=-\beta _{1}/3$ two \textit{arbitrary}
parameters, is \textit{solvable}: indeed the solution of its initial-values
problem---to evaluate $x_{1}\left( t\right) $ and $x_{2}\left( t\right) $
from \textit{arbitrarily} assigned initial values $x_{1}\left( 0\right) $
and $x_{2}\left( 0\right) $---are (\textit{explicitly}!) yielded by the
solution of a \textit{quadratic} algebraic equation the coefficients of
which involve the Jacobian elliptic function $\mu $ sn$\left( \lambda t+\rho
,k\right) $ with the $4$ parameters $\mu ,$ $\lambda ,$ $\rho ,$ $k$ given
by \textit{simple} formulas in terms of the $2$ initial data $x_{1}\left(
0\right) $ and $x_{2}\left( 0\right) $ and the $2$ a \textit{priori arbitrary%
} parameters $a$ and $b$. (The interested reader can easily obtain all the
relevant formulas from the treatment given above, comparing it if need be
with the analogous treatment provided in \textbf{Example 1 }of \cite{CP2018}%
; or see below \textbf{Subsection 4.7)}.

\textbf{Remark 3.1-2}. An equivalent---indeed more direct---way to identify
the \textit{solvable} dynamical system (\ref{311x1x2dot}) as corresponding
to the \textit{solvable} dynamical system
\end{subequations}
\begin{equation}
\dot{y}_{1}=\frac{4}{3}\beta _{0}-\frac{32\beta _{1}}{9}y_{2}~,~~~\dot{y}%
_{2}=\beta _{0}y_{1}+\beta _{1}\left( y_{1}\right) ^{3}
\end{equation}%
(see (\ref{311y1y2dot}) and (\ref{M4N2f12})), is via the relations
\begin{subequations}
\begin{equation}
y_{1}=-\left( 3x_{1}+x_{2}\right) ~,~~~y_{2}=3x_{1}\left( x_{1}+x_{2}\right)
\end{equation}%
(see (\ref{M4N2mu31ym})) and their time derivatives,%
\begin{equation}
\dot{y}_{1}=-\left( 3\dot{x}_{1}+\dot{x}_{2}\right) ~,~~~\dot{y}_{2}=3\left[
\left( 2x_{1}+x_{2}\right) \dot{x}_{1}+x_{1}\dot{x}_{2}\right]
~.~\blacksquare
\end{equation}

\bigskip

\subsection{Case (ii): $\protect\mu _{1}=\protect\mu _{2}=2$}

In this case
\end{subequations}
\begin{subequations}
\begin{eqnarray}
y_{1} &=&-2\left( x_{1}+x_{2}\right) ~,~~~y_{2}=\left( x_{1}\right)
^{2}+\left( x_{2}\right) ^{2}+4x_{1}x_{2}~,  \notag \\
y_{3} &=&-2x_{1}x_{2}\left( x_{1}+x_{2}\right) ~,~~~y_{4}=\left(
x_{1}x_{2}\right) ^{2}~.  \label{M4N2mu22ym}
\end{eqnarray}

\textbf{Remark 3.2-1}. Of course a remark completely analogous to \textbf{%
Remark 3.1-1} holds in this case as well. $\blacksquare $

The corresponding equations (\ref{2Eqydot0}) read
\begin{equation}
\dot{y}_{1}\left( x_{n}\right) ^{3}+\dot{y}_{2}\left( x_{n}\right) ^{2}+\dot{%
y}_{3}x_{n}+\dot{y}_{4}=0~,~~~n=1,2~;
\end{equation}%
and proceeding as above one easily obtains, for the $6$ assignments (\ref%
{M4N2mu13mtilde}), the following systems of $2$ ODEs:
\end{subequations}
\begin{subequations}
\begin{equation}
\dot{x}_{1}=-\frac{\left( 2x_{1}+x_{2}\right) \dot{y}_{1}+\dot{y}_{2}}{%
2\left( x_{1}-x_{2}\right) }~,~~~\dot{x}_{2}=\frac{\left(
x_{1}+2x_{2}\right) \dot{y}_{1}+\dot{y}_{2}}{2\left( x_{1}-x_{2}\right) }~,
\label{M4N2mu22x12dot}
\end{equation}%
\begin{equation}
\dot{x}_{1}=-\frac{x_{1}\left( x_{1}+2x_{2}\right) \dot{y}_{1}-\dot{y}_{3}}{2%
\left[ \left( x_{1}\right) ^{2}-\left( x_{2}\right) ^{2}\right] }~,~~~\dot{x}%
_{2}=\frac{x_{2}\left( x_{2}+2x_{1}\right) \dot{y}_{1}+\dot{y}_{3}}{2\left[
\left( x_{1}\right) ^{2}-\left( x_{2}\right) ^{2}\right] }~,
\end{equation}%
\begin{equation}
\dot{x}_{1}=-\frac{\left( x_{1}\right) ^{2}x_{2}\dot{y}_{1}+\dot{y}_{4}}{%
2x_{1}x_{2}\left( x_{1}-x_{2}\right) }~,~~~\dot{x}_{2}=\frac{x_{1}\left(
x_{2}\right) ^{2}\dot{y}_{1}+\dot{y}_{4}}{2x_{1}x_{2}\left(
x_{1}-x_{2}\right) }~,
\end{equation}%
\begin{equation}
\dot{x}_{1}=\frac{x_{1}\left( x_{1}+2x_{2}\right) \dot{y}_{2}+\left(
2x_{1}+x_{2}\right) \dot{y}_{3}}{2\left[ \left( x_{1}\right) ^{3}-\left(
x_{2}\right) ^{3}\right] }~,~~~\dot{x}_{2}=-\frac{x_{2}\left(
x_{2}+2x_{1}\right) \dot{y}_{2}+\left( 2x_{2}+x_{1}\right) \dot{y}_{3}}{2%
\left[ \left( x_{1}\right) ^{3}-\left( x_{2}\right) ^{3}\right] }~,
\end{equation}%
\begin{equation}
\dot{x}_{1}=\frac{\left( x_{1}\right) ^{2}x_{2}\dot{y}_{2}-\left(
2x_{1}+x_{2}\right) \dot{y}_{4}}{2x_{1}x_{2}\left[ \left( x_{1}\right)
^{2}-\left( x_{2}\right) ^{2}\right] }~,~~~\dot{x}_{2}=-\frac{\left(
x_{2}\right) ^{2}x_{1}\dot{y}_{2}-\left( 2x_{2}+x_{1}\right) \dot{y}_{4}}{%
2x_{1}x_{2}\left[ \left( x_{1}\right) ^{2}-\left( x_{2}\right) ^{2}\right] }
\end{equation}

\begin{equation}
\dot{x}_{1}=-\frac{x_{1}x_{2}\dot{y}_{3}+\left( x_{1}+2x_{2}\right) \dot{y}%
_{4}}{2x_{1}\left( x_{2}\right) ^{2}\left( x_{1}-x_{2}\right) }~,~~~\dot{x}%
_{2}=\frac{x_{1}x_{2}\dot{y}_{3}+\left( x_{2}+2x_{1}\right) \dot{y}_{4}}{%
2x_{2}\left( x_{1}\right) ^{2}\left( x_{1}-x_{2}\right) }~.
\end{equation}

Hence, to the system (\ref{311y1y2dot}), one now associates again the
requirement (\ref{31ConPol}); and---by making again the assignment (\ref%
{M4N2f12a}) for the system of evolution equations satisfied by $y_{1}\left(
t\right) $ and $y_{2}\left( t\right) $---one identifies again the
restriction (\ref{M4N2falphabeta}), thereby concluding---via (\ref%
{M4N2mu22x12dot})---that the polynomial system
\end{subequations}
\begin{equation}
\dot{x}_{n}=a+b\left[ \left( x_{n}\right) ^{2}-8x_{n}x_{n+1}-5\left(
x_{n+1}\right) ^{2}\right] ~,~~~n=1,2~~\func{mod}\left[ 2\right] ~,
\label{312x1x2dot}
\end{equation}%
where now $a=-\beta _{0}/3$ and $b=4\beta _{1}/9$, is \textit{solvable}. And
the \textit{explicit} solution is then quite analogous (up to simple
modifications of some parameters) to that described (after eq. (\ref%
{311x1x2dot})) in the preceding \textbf{Subsection 3.1}.

\bigskip

\section{Other \textit{solvable} systems of 2 nonlinearly-coupled ODEs
identified via the technique described in Section 3}

In this \textbf{Section 4} we report a list of \textit{solvable} systems of $%
2$ \textit{nonlinearly coupled first-order} ODEs satisfied by the $2$
dependent variables $x_{1}\left( t\right) $ and $x_{2}\left( t\right) $; in
each case we identify the corresponding \textit{solvable }system of $2$ ODEs
satisfied by $2$ variables $y_{\tilde{m}}\left( t\right) $ (for these, and
other, notations used below see \textbf{Section 3}); indeed, to help the
reader mainly interested in the \textit{solvable} character of one of the
following systems we also specify below on a case-by-case basis the
information which allows to \textit{solve} that specific system (we do so
even at the cost of minor repetitions). Note that the majority of these
models feature equations of type (\ref{1xndotPol}), but in a few cases the
right-hand sides of these ODEs are \textit{not quite polynomial}. And let us
recall that in this \textbf{Section 4} parameters such as $a,$ $b,$ $c$
(possibly equipped with indices) are \textit{arbitrary} numbers (possibly
\textit{complex}).

\textbf{Remark 4-1}. Most of the models reported below are characterized by
evolution equations of the following kind:
\begin{subequations}
\begin{equation}
\dot{x}_{n}=\sum_{k=0}^{K}\left[ p_{k}^{\left( n\right) }\left(
x_{1},x_{2}\right) \right] ~,~~~n=1,2~,
\end{equation}%
with $K$ a \textit{positive integer} and the functions $p_{k}^{\left(
n\right) }\left( x_{1},x_{2}\right) $ \textit{homogenous polynomials of
degree} $k$,%
\begin{equation}
p_{k}^{\left( n\right) }\left( x_{1},x_{2}\right) =\sum_{\ell =0}^{k}\left[
a_{\ell }^{\left( n,k\right) }\left( x_{1}\right) ^{k-\ell }\left(
x_{2}\right) ^{\ell }\right] ~,~~~k=0,1,...,K~,~~~n=1,2~.
\end{equation}%
So the different models are characterized by the assignments of the \textit{%
positive integer} $K$ and of the $\left( K+1\right) ^{2}$ parameters $%
a_{\ell }^{\left( n,k\right) }$, expressed in each case in terms of a
\textit{few arbitrary parameters}. It is of course obvious that in all the
models associated with \textbf{Case (ii)} (see \textbf{Subsection 3.2})
these parameters satisfy the restriction $a_{\ell }^{\left( 1,k\right)
}=a_{\ell }^{\left( 2,k\right) }$, since in that case the $2$ zeros $%
x_{1}\left( t\right) $ and $x_{2}\left( t\right) $ are completely
equivalent; while this restriction \textit{need not} hold in \textbf{Case (i)%
} (see \textbf{Subsection 3.1}), although in some such cases it also emerges
(see below). It is on the other hand plain that, \textit{also} in \textbf{%
Case (i)} (as, obviously, in \textbf{Case (ii)}), there holds the restriction%
\begin{equation}
\sum_{\ell =0}^{k}\left[ a_{\ell }^{\left( 1,k\right) }\right] =\sum_{\ell
=0}^{k}\left[ a_{\ell }^{\left( 2,k\right) }\right] ~,~~~k=0,1,...,K~,
\end{equation}%
because for the special initial conditions $x_{1}\left( 0\right)
=x_{2}\left( 0\right) $---implying $x_{1}\left( t\right) =x_{2}\left(
t\right) \equiv x\left( t\right) $, since in such case the distinction among
\textbf{Case (i)} and \textbf{Case (ii)} obviously disappears---the $2$
evolution equations (with $n=1,2$) satisfied by $x\left( t\right) $,
\begin{equation}
\dot{x}=x^{k}\sum_{\ell =0}^{k}\left[ a_{\ell }^{\left( n,k\right) }\right]
~,~~~k=0,1,...,K~,~~~n=1,2~,
\end{equation}%
must coincide. $\blacksquare $

\textbf{Remark 4-2}. In the following $44$ subsections we list as many
\textit{solvable} systems of $2$ nonlinearly-coupled first-order ODEs, most
of them with polynomial right-hand sides, and we indicate how each of them
can be solved. The presentation of all these models is made so as to
facilitate the utilization of these findings by practitioners only
interested in one of these models (or its generalization, see \textbf{%
Section 5}). Note however that not all these models are different among
themselves: indeed, some feature \textit{identical} equations of
motion---although the method to solve them might seem different. This is
demonstrated by the following self-evident identification of the following
equations of motion: (\ref{4i12c})$\equiv $(\ref{4i23b}), (\ref{4ii12c})$%
\equiv $(\ref{4ii23b}), (\ref{4i13c})$\equiv $(\ref{4i23cc})$\equiv $(\ref%
{4i34c}), (\ref{4ii13c})$\equiv $(\ref{4ii23c})$\equiv $(\ref{4ii34cc}), (%
\ref{4i14c})$\equiv $(\ref{4i24c})$\equiv $(\ref{4i34b}), (\ref{4ii14c})$%
\equiv $(\ref{4ii24c})$\equiv $(\ref{4ii34b}). So in fact the list below
contains \textit{only} $34$ \textit{different} systems of $2$
nonlinearly-coupled first-order differential equations for the $2$
time-dependent variables $x_{1}\left( t\right) $ and $x_{2}\left( t\right) $%
. $\blacksquare $

\bigskip

\subsection{Model 4.(i)1.2a}

\end{subequations}
\begin{subequations}
\begin{equation}
\dot{x}_{1}=x_{1}\left\{ a+b\left[ 11\left( x_{1}\right)
^{2}+6x_{1}x_{2}-\left( x_{2}\right) ^{2}\right] \right\} ~,
\end{equation}%
\begin{equation}
\dot{x}_{2}=ax_{2}+b\left[ -6\left( x_{1}\right) ^{3}+9\left( x_{1}\right)
^{2}x_{2}+12x_{1}\left( x_{2}\right) ^{2}+\left( x_{2}\right) ^{3}\right] ~;
\end{equation}%
$x_{1}\left( t\right) $ and $x_{2}\left( t\right) $ are related to $%
y_{1}\left( t\right) $ and $y_{2}\left( t\right) $ by (\ref{M4N2mu31ym});
and the variables $y_{1}\left( t\right) $ and $y_{2}\left( t\right) $ evolve
according to (\ref{A1SolvSyst}), the \textit{explicit} solution of which is
given by the relevant formulas in \textbf{Subsection Case A.1} of \textbf{%
Appendix A} with $\tilde{m}_{1}=1,$ $\tilde{m}_{2}=2,$ $L=1,$ $\alpha
_{0}=a, $ $\alpha _{1}=3b,$ $\beta _{0}=2a,$ $\beta _{1}=16b.$

\bigskip

\subsection{Model 4.(ii)1.2a}

\end{subequations}
\begin{equation}
\dot{x}_{n}=ax_{n}+b\left[ 4\left( x_{n}\right) ^{3}+9\left( x_{n}\right)
^{2}x_{n+1}-\left( x_{n+1}\right) ^{3}\right] ~,~\ ~n=1,2~\func{mod}\left[ 2%
\right] ~;
\end{equation}%
$x_{1}\left( t\right) $ and $x_{2}\left( t\right) $ are related to $%
y_{1}\left( t\right) $ and $y_{2}\left( t\right) $ by (\ref{M4N2mu22ym});
and the variables $y_{1}\left( t\right) $ and $y_{2}\left( t\right) $ evolve
according to (\ref{A1SolvSyst}), the \textit{explicit} solution of which is
given by the relevant formulas in \textbf{Subsection} \textbf{Case A.1} of
\textbf{Appendix A} with $\tilde{m}_{1}=1,$ $\tilde{m}_{2}=2,$ $L=1,$ $%
\alpha _{0}=a,$ $\alpha _{1}=\left( 3/4\right) b,$ $\beta _{0}=2a,$ $\beta
_{1}=4b.$

\bigskip

\subsection{Model 4.(i)1.2b}

\begin{subequations}
\begin{eqnarray}
\dot{x}_{n} &=&a_{0}+x_{n}\left( a_{1}+a_{2}X+a_{3}X^{2}\right)
+b_{1}X+b_{2}X^{2}+b_{3}X^{3}~,  \notag \\
X &\equiv &3x_{1}+x_{2},~~~n=1,2~;  \label{4i12bb}
\end{eqnarray}%
$x_{1}\left( t\right) $ and $x_{2}\left( t\right) $ are related to $%
y_{1}\left( t\right) $ and $y_{2}\left( t\right) $ by (\ref{M4N2mu31ym});
and the variables $y_{1}\left( t\right) $ and $y_{2}\left( t\right) $ evolve
according to (\ref{A2SolvSyst}), the \textit{explicit} solution of which is
given by the relevant formulas in \textbf{Subsection} \textbf{Case A.2} of
\textbf{Appendix A} with $\tilde{m}_{1}=1,$ $\tilde{m}_{2}=2,$ $L=3,$ $%
\alpha _{0}=-4a_{0},$ $\alpha _{1}=a_{1}+4b_{1},$ $\alpha _{2}=-\left(
a_{2}+4b_{2}\right) ,$ $\alpha _{3}=a_{3}+4b_{3},$ $\beta _{1}=2a_{1},$ $%
\beta _{2}=-2a_{2},$ $\beta _{3}=2a_{3},$ $\gamma _{0}=-3a_{0},$ $\gamma
_{1}=3b_{1},$ $\gamma _{2}=-3b_{2},$ $\gamma _{3}=3b_{3}.$

This model (\ref{4i12bb}) (with $a_{0}=0$) is actually a special case of the
more general model%
\begin{equation}
\dot{x}_{n}=a_{0}+\sum_{\ell =1}^{L}\left\{ \left( -X\right) ^{\ell -1}\left[
a_{\ell }x_{n}+b_{\ell }X\right] \right\} ~,~~~X\equiv
3x_{1}+x_{2},~~~n=1,2~,
\end{equation}%
again with $x_{1}\left( t\right) $ and $x_{2}\left( t\right) $ related to $%
y_{1}\left( t\right) $ and $y_{2}\left( t\right) $ by (\ref{M4N2mu31ym}) and
the variables $y_{1}\left( t\right) $ and $y_{2}\left( t\right) $ evolving
according to (\ref{A2SolvSyst}) with $\tilde{m}_{1}=1,$ $\tilde{m}_{2}=2,$ $%
L $ an \textit{arbitrary} positive integer, $\alpha _{0}=-4a_{0},$ $\alpha
_{\ell }=a_{\ell }+4b_{\ell },$ $\beta _{\ell }=2a_{\ell },$ $\gamma
_{0}=-3a_{0},$ $\gamma _{\ell }=3b_{\ell }.$

\bigskip

\subsection{Model 4.(ii)1.2b}

\end{subequations}
\begin{subequations}
\begin{eqnarray}
&&\dot{x}_{n}=a_{0}+x_{n}\left( a_{1}+a_{2}X+a_{3}X^{2}\right)
+b_{1}X+b_{2}X^{2}+b_{3}X^{3}~,  \notag \\
&&X\equiv x_{1}+x_{2}~,~~~n=1,2~;  \label{4ii12b}
\end{eqnarray}%
$x_{1}\left( t\right) $ and $x_{2}\left( t\right) $ are related to $%
y_{1}\left( t\right) $ and $y_{2}\left( t\right) $ by (\ref{M4N2mu22ym});
and the variables $y_{1}\left( t\right) $ and $y_{2}\left( t\right) $ evolve
according to (\ref{A2SolvSyst}), the \textit{explicit} solution of which is
given by the relevant formulas in \textbf{Subsection Case A.2} of \textbf{%
Appendix A} with $\tilde{m}_{1}=1,$ $\tilde{m}_{2}=2,$ $L=3,$ $\alpha
_{0}=-4a_{0},$ $\alpha _{1}=a_{1}+2b_{1},$ $\alpha _{2}=-a_{2}/2-b_{2},$ $%
\alpha _{3}=b_{3}/2+a_{3}/4,$ $\beta _{1}=2a_{1},$ $\beta _{2}=-a_{2},$ $%
\beta _{3}=a_{3}/2,$ $\gamma _{0}=-3a_{0},$ $\gamma _{1}=\left( 3/2\right)
b_{1},$ $\gamma _{2}=-\left( 3/4\right) b_{2},$ $\gamma _{3}=\left(
3/8\right) b_{3}.$

This model (\ref{4ii12b}) is actually a special case of the more general
model%
\begin{equation}
\dot{x}_{n}=a_{0}+\sum_{\ell =1}^{L}\left\{ \left( -2X\right) ^{\ell -1}%
\left[ a_{\ell }x_{n}+b_{\ell }X\right] \right\} ~,~~~X\equiv
x_{1}+x_{2}~,~~~n=1,2~~\func{mod}\left[ 2\right] ~,
\end{equation}%
again with $x_{1}\left( t\right) $ and $x_{2}\left( t\right) $ related to $%
y_{1}\left( t\right) $ and $y_{2}\left( t\right) $ by (\ref{M4N2mu22ym}) and
the variables $y_{1}\left( t\right) $ and $y_{2}\left( t\right) $ evolving
according to (\ref{A2SolvSyst}) with $\tilde{m}_{1}=1,$ $\tilde{m}_{2}=2,$ $%
L $ an \textit{arbitrary} positive integer, $\alpha _{0}=-4a_{0},$ $\alpha
_{\ell }=a_{\ell }+2b_{\ell },$ $\beta _{\ell }=2a_{\ell },$ $\gamma
_{0}=-3a_{0},$ $\gamma _{\ell }=\left( 3/2\right) b_{\ell }.$

\bigskip

\subsection{Model 4.(i)1.2c}

\end{subequations}
\begin{equation}
\dot{x}_{n}=x_{n}\left[ a+bX+cX^{2}\right] ~,~~~X\equiv x_{1}\left(
x_{1}+x_{2}\right) ~,~~\hspace{0in}n=1,2~;  \label{4i12c}
\end{equation}%
$x_{1}\left( t\right) $ and $x_{2}\left( t\right) $ are related to $%
y_{1}\left( t\right) $ and $y_{2}\left( t\right) $ by (\ref{M4N2mu31ym});
and the variables $y_{1}\left( t\right) $ and $y_{2}\left( t\right) $ evolve
according to (\ref{A2SolvSyst}), the \textit{explicit} solution of which is
given by the relevant formulas in \textbf{Subsection} \textbf{Case A.2} of
\textbf{Appendix A} with $\tilde{m}_{1}=2,$ $\tilde{m}_{2}=1,$ $L=3,$ $%
\alpha _{0}=0,$ $\alpha _{1}=2a,$ $\alpha _{2}=2b/3,$ $\alpha _{3}=2c/9,$ $%
\beta _{1}=a,$ $\beta _{2}=b/3,$ $\beta _{3}=c/9,$ $\gamma _{\ell }=0.$

\bigskip

\subsection{Model 4.(ii)1.2c}

\begin{eqnarray}
\dot{x}_{n} &=&x_{n}\left( a+bX+cX^{2}\right) ~,  \notag \\
X &\equiv &\left( x_{1}\right) ^{2}+4x_{1}x_{2}+\left( x_{2}\right)
^{2}~,~~~n=1,2~;  \label{4ii12c}
\end{eqnarray}%
$x_{1}\left( t\right) $ and $x_{2}\left( t\right) $ are related to $%
y_{1}\left( t\right) $ and $y_{2}\left( t\right) $ by (\ref{M4N2mu22ym});
and the variables $y_{1}\left( t\right) $ and $y_{2}\left( t\right) $ evolve
according to (\ref{A2SolvSyst}), the \textit{explicit} solution of which is
given by the relevant formulas in \textbf{Subsection} \textbf{Case A.2} of
\textbf{Appendix A} with $\tilde{m}_{1}=2,$ $\tilde{m}_{2}=1,$ $L=3,$ $%
\alpha _{0}=0,$ $\alpha _{1}=2a,$ $\alpha _{2}=2b,$ $\alpha _{3}=2c,$ $\beta
_{1}=a,$ $\beta _{2}=b,$ $\beta _{3}=c,$ $\gamma _{\ell }=0.$

\bigskip

\subsection{Model 4.(i)1.2d}

\begin{subequations}
\label{4i12d}
\begin{equation}
\dot{x}_{1}=a+b\left[ 5\left( x_{1}\right) ^{2}+10x_{1}x_{2}+\left(
x_{2}\right) ^{2}\right] ~,  \label{4i12da}
\end{equation}%
\begin{equation}
\dot{x}_{2}=a+b\left[ 17\left( x_{1}\right) ^{2}+2x_{1}x_{2}-3\left(
x_{2}\right) ^{2}\right] ~;  \label{4i12db}
\end{equation}%
$x_{1}\left( t\right) $ and $x_{2}\left( t\right) $ are related to $%
y_{1}\left( t\right) $ and $y_{2}\left( t\right) $ by (\ref{M4N2mu31ym});
and the variables $y_{1}\left( t\right) $ and $y_{2}\left( t\right) $ evolve
according to (\ref{A3SolvSysta}), the \textit{explicit} solution of which is
given by the relevant formulas in \textbf{Subsection Case A.3.1} of \textbf{%
Appendix A} with $\tilde{m}_{1}=1,$ $\tilde{m}_{2}=2,$ $\alpha _{0}=-4a,$ $%
\alpha _{1}=-\left( 32/3\right) b,$ $\beta _{0}=-3a,$ $\beta _{1}=-3b$. Note
that this is the model treated in detail in \textbf{Subsection 3.1.1}, see (%
\ref{311x1x2dot}).

\bigskip

\subsection{Model 4.(ii)1.2d}

\end{subequations}
\begin{equation}
\dot{x}_{n}=a+b\left[ \left( x_{n}\right) ^{2}-8x_{1}x_{2}-5\left(
x_{n+1}\right) ^{2}\right] ~,~\ ~n=1,2~\func{mod}\left[ 2\right] ~;
\end{equation}%
$x_{1}\left( t\right) $ and $x_{2}\left( t\right) $ are related to $%
y_{1}\left( t\right) $ and $y_{2}\left( t\right) $ by (\ref{M4N2mu22ym});
and the variables $y_{1}\left( t\right) $ and $y_{2}\left( t\right) $ evolve
according to (\ref{A3SolvSysta}), the \textit{explicit} solution of which is
given by the relevant formulas in \textbf{Subsection Case A.3.1} of \textbf{%
Appendix A} with $\tilde{m}_{1}=1,$ $\tilde{m}_{2}=2,$ $\alpha _{0}=-4a,$ $%
\alpha _{1}=8b,$ $\beta _{0}=-3a,$ $\beta _{1}=\left( 9/4\right) b$. Note
that this is the model treated in detail in \textbf{Subsection 3.1.2}, see (%
\ref{312x1x2dot}).

\bigskip

\subsection{Model 4.(i)1.3a}

\begin{subequations}
\begin{equation}
\dot{x}_{1}=x_{1}\left\{ a+b\left[ 65\left( x_{1}\right) ^{3}+77\left(
x_{1}\right) ^{2}x_{2}-13x_{1}\left( x_{2}\right) ^{2}-\left( x_{2}\right)
^{3}\right] \right\} ~,
\end{equation}%
\begin{equation}
\dot{x}_{2}=ax_{2}-b\left[ 33\left( x_{1}\right) ^{4}+15\left( x_{1}\right)
^{3}x_{2}-147\left( x_{1}\right) ^{2}\left( x_{2}\right) ^{2}-27x_{1}\left(
x_{2}\right) ^{3}-2\left( x_{2}\right) ^{4}\right] ~;
\end{equation}%
$x_{1}\left( t\right) $ and $x_{2}\left( t\right) $ are related to $%
y_{1}\left( t\right) $ and $y_{3}\left( t\right) $ by (\ref{M4N2mu31ym});
and the variables $y_{1}\left( t\right) $ and $y_{3}\left( t\right) $ evolve
according to (\ref{A1SolvSyst}), the \textit{explicit} solution of which is
given by the relevant formulas in \textbf{Subsection} \textbf{Case A.1} of
\textbf{Appendix A}) with $\tilde{m}_{1}=1,$ $\tilde{m}_{2}=3,$ $L=1,$ $%
\alpha _{0}=a,$ $\alpha _{1}=-2b,$ $\beta _{0}=3a,$ $\beta _{1}=-96b.$

\bigskip

\subsection{Model 4.(ii)1.3a}

\end{subequations}
\begin{equation}
\dot{x}_{n}=x_{n}\left\{ a+b\left( x_{1}+x_{2}\right) \left[ \left(
x_{n}\right) ^{2}+5x_{1}x_{2}-2\left( x_{n+1}\right) ^{2}\right] \right\}
~,~~~n=1,2~~\func{mod}\left[ 2\right] ~;
\end{equation}%
$x_{1}\left( t\right) $ and $x_{2}\left( t\right) $ are related to $%
y_{1}\left( t\right) $ and $y_{3}\left( t\right) $ by (\ref{M4N2mu22ym});
and the variables $y_{1}\left( t\right) $ and $y_{3}\left( t\right) $ evolve
according to (\ref{A1SolvSyst}), the \textit{explicit} solution of which is
given by the relevant formulas in \textbf{Subsection Case A.1} of \textbf{%
Appendix A} with $\tilde{m}_{1}=1,$ $\tilde{m}_{2}=3,$ $L=1,$ $\alpha
_{0}=a, $ $\alpha _{1}=-b/8,$ $\beta _{0}=3a,$ $\beta _{1}=-6b.$

\bigskip

\subsection{Model 4.(i)1.3b}

\begin{subequations}
\label{4i13b}
\begin{eqnarray}
&&\dot{x}_{1}=\left( 6x_{1}\right) ^{-1}\left[ \left( x_{1}\right)
^{2}\left( a_{0}+a_{1}X+a_{2}X^{2}\right) \right.  \notag \\
&&\left. +\left( 7x_{1}+x_{2}\right) \left(
b_{0}+b_{1}X+b_{2}X^{2}+b_{3}X^{3}\right) \right] ~,
\end{eqnarray}%
\begin{eqnarray}
\dot{x}_{2} &=&\left( 6x_{1}\right) ^{-1}\left[ x_{1}x_{2}\left(
a_{0}+a_{1}X+a_{2}X^{2}\right) \right.  \notag \\
&&\left. +\left( 11x_{1}-3x_{2}\right) \left(
b_{0}+b_{1}X+b_{2}X^{2}+b_{3}X^{3}\right) \right] ~,
\end{eqnarray}%
\begin{equation}
X\equiv 3x_{1}+x_{2}~;
\end{equation}%
$x_{1}\left( t\right) $ and $x_{2}\left( t\right) $ are related to $%
y_{1}\left( t\right) $ and $y_{3}\left( t\right) $ by (\ref{M4N2mu31ym});
and the variables $y_{1}\left( t\right) $ and $y_{3}\left( t\right) $ evolve
according to (\ref{A2SolvSyst}), the \textit{explicit} solution of which is
given by the relevant formulas in \textbf{Subsection} \textbf{Case A.2} of
\textbf{Appendix A} with $\tilde{m}_{1}=1,$ $\tilde{m}_{2}=3,$ $L=3,$ $%
\alpha _{0}=-\left( 16/3\right) b_{0},$ $\alpha _{\ell }=\left( -1\right)
^{\ell -1}\left( a_{\ell -1}/6\right) +\left( 16/3\right) b_{\ell },$ $\beta
_{\ell }=\left( -1\right) ^{\ell -1}a_{\ell -1}/2,$ $\left( \ell
=1,2,3\right) ,~\gamma _{\ell -1}=\left( -1\right) ^{\ell }b_{\ell -1},$ $%
\left( \ell =1,2,3,4\right) .$ Note that the right-hand sides of these $2$
ODEs, (\ref{4i13b}), are both \textit{polynomial} only if the $4$ parameters
$b_{\ell }$ vanish, $b_{\ell }=0,$ $\ell =0,1,2,3$.

\bigskip

\subsection{Model 4.(ii)1.3b}

\end{subequations}
\begin{eqnarray}
&&\dot{x}_{1}=6^{-1}\left[ x_{n}\left( a_{1}+a_{2}X+a_{3}X^{2}\right) \right.
\notag \\
&&\left. +\left( x_{n}+3x_{n+1}\right) \left(
b_{0}X^{-1}+b_{1}+b_{2}X+b_{3}X^{2}\right) \right] ~,  \notag \\
&&X\equiv x_{1}+x_{2}~,~~~n=1,2~~\func{mod}\left[ 2\right] ~;
\end{eqnarray}%
$x_{1}\left( t\right) $ and $x_{2}\left( t\right) $ are related to $%
y_{1}\left( t\right) $ and $y_{3}\left( t\right) $ by (\ref{M4N2mu22ym});
and the variables $y_{1}\left( t\right) $ and $y_{3}\left( t\right) $ evolve
according to (\ref{A2SolvSyst}), the \textit{explicit} solution of which is
given by the relevant formulas in \textbf{Subsection} \textbf{Case A.2} of
\textbf{Appendix A} with $\tilde{m}_{1}=1,$ $\tilde{m}_{2}=3,$ $L=3,$ $%
\alpha _{0}=-\left( 16/3\right) b_{0},$ $\alpha _{\ell }=-\left( -2\right)
^{-\ell }\left( a_{\ell }+2^{5-\ell }b_{\ell }\right) /3,$ $\beta _{\ell
}=-\left( -2\right) ^{-\ell -1}a_{\ell },$ $\gamma _{0}=-b_{0},$ $\gamma
_{\ell }=-\left( -2\right) ^{-\ell }b_{\ell },$ $\ell =1,2,3.$ Note that the
right-hand sides of these $2$ ODEs, (\ref{4i13b}), are both \textit{%
polynomial} iff the single parameter $b_{0}$ vanishes, $b_{0}=0$.

\bigskip

\subsection{Model 4.(i)1.3c}

\begin{equation}
\dot{x}_{n}=x_{n}\left( a+bX+cX^{2}\right) ~,~~~X\equiv \left( x_{1}\right)
^{2}\left( x_{1}+3x_{2}\right) ~,~~~n=1,2~;  \label{4i13c}
\end{equation}%
$x_{1}\left( t\right) $ and $x_{2}\left( t\right) $ are related to $%
y_{1}\left( t\right) $ and $y_{3}\left( t\right) $ by (\ref{M4N2mu31ym});
and the variables $y_{1}\left( t\right) $ and $y_{3}\left( t\right) $ evolve
according to (\ref{A2SolvSyst}), the \textit{explicit} solution of which is
given by the relevant formulas in \textbf{Subsection Case A.2} of \textbf{%
Appendix A} with $\tilde{m}_{1}=3,$ $\tilde{m}_{2}=1,$ $L=3,$ $\alpha
_{0}=0, $ $\alpha _{1}=3a,$ $\alpha _{2}=-3b,$ $\alpha _{3}=3c,$ $\beta
_{1}=a,$ $\beta _{2}=-b,$ $\beta _{3}=c,$ $\gamma _{\ell }=0.$

\bigskip

\subsection{Model 4.(ii)1.3c}

\begin{equation}
\dot{x}_{n}=x_{n}\left( a+bX+cX^{2}\right) ~,X\equiv x_{1}x_{2}\left(
x_{1}+x_{2}\right) ~,~~~n=1,2~;  \label{4ii13c}
\end{equation}%
$x_{1}\left( t\right) $ and $x_{2}\left( t\right) $ are related to $%
y_{1}\left( t\right) $ and $y_{3}\left( t\right) $ by (\ref{M4N2mu22ym});
and the variables $y_{1}\left( t\right) $ and $y_{3}\left( t\right) $ evolve
according to (\ref{A2SolvSyst}), the \textit{explicit} solution of which is
given by the relevant formulas in \textbf{Subsection Case A.2} of \textbf{%
Appendix A}) with $\tilde{m}_{1}=3,$ $\tilde{m}_{2}=1,$ $L=3,$ $\alpha
_{0}=0,$ $\alpha _{1}=3a,$ $\alpha _{2}=-3b/2,$ $\alpha _{3}=3c/4,$ $\beta
_{1}=a,$ $\beta _{2}=-b/2,$ $\beta _{3}=c/4,$ $\gamma _{\ell }=0.$

\bigskip

\subsection{Model 4.(i)1.3d}

\begin{subequations}
\label{4i13d}
\begin{eqnarray}
\dot{x}_{1}=\left( 6x_{1}\right) ^{-1}\left\{ a\left( 7x_{1}+x_{2}\right)
\right. &&  \notag \\
\left. +b\left[ 13\left( x_{1}\right) ^{4}+376\left( x_{1}\right)
^{3}x_{2}+106\left( x_{1}x_{2}\right) ^{2}+16x_{1}\left( x_{2}\right)
^{3}+\left( x_{2}\right) ^{4}\right] \right\} ~, &&
\end{eqnarray}%
\begin{eqnarray}
\dot{x}_{2}=\left( 6x_{1}\right) ^{-1}\left\{ a\left( 11x_{1}-3x_{2}\right)
\right. &&  \notag \\
\left. +b\left[ 473\left( x_{1}\right) ^{4}+408\left( x_{1}\right)
^{3}x_{2}-318\left( x_{1}x_{2}\right) ^{2}-48x_{1}\left( x_{2}\right)
^{3}-3\left( x_{2}\right) ^{4}\right] \right\} ~; &&
\end{eqnarray}%
$x_{1}\left( t\right) $ and $x_{2}\left( t\right) $ are related to $%
y_{1}\left( t\right) $ and $y_{3}\left( t\right) $ by (\ref{M4N2mu31ym});
and the variables $y_{1}\left( t\right) $ and $y_{3}\left( t\right) $ evolve
according to (\ref{A3SolvSysta}), the \textit{explicit} solution of which is
given by the relevant formulas in \textbf{Subsection Case A.3.2 }of \textbf{%
Appendix A} with $\tilde{m}_{1}=1,$ $\tilde{m}_{2}=3,$ $\alpha _{0}=-\left(
16/3\right) a,$ $\alpha _{1}=\left( 16^{2}/3\right) b,$ $\beta _{0}=-a,$ $%
\beta _{1}=b$. Note that the right-hand sides of the $2$ ODEs (\ref{4i13d})
are \textit{not} polynomial.

\bigskip

\subsection{Model 4.(ii)1.3d}

\end{subequations}
\begin{eqnarray}
&&\dot{x}_{1}=a\left( \frac{x_{n}+3x_{n+1}}{x_{1}+x_{2}}\right)  \notag \\
&&+b\left[ 3\left( x_{n}\right) ^{3}-\left( x_{n}\right)
^{2}x_{n+1}-~15x_{n}\left( x_{n+1}\right) ^{2}-3\left( x_{n+1}\right) ^{3}%
\right] ~,  \notag \\
&&n=1,2~~\func{mod}\left[ 2\right] ~;  \label{4ii13d}
\end{eqnarray}%
$x_{1}\left( t\right) $ and $x_{2}\left( t\right) $ are related to $%
y_{1}\left( t\right) $ and $y_{3}\left( t\right) $ by (\ref{M4N2mu22ym});
and the variables $y_{1}\left( t\right) $ and $y_{3}\left( t\right) $ evolve
according to (\ref{A3SolvSysta}), the \textit{explicit} solution of which is
given by the relevant formulas in \textbf{Subsection Case A.3.2 }of \textbf{%
Appendix A} with $\tilde{m}_{1}=1,$ $\tilde{m}_{2}=3,$ $\alpha _{0}=-8a,$ $%
\alpha _{1}=-16b,$ $\beta _{0}=-\left( 3/2\right) a,$ $\beta _{1}=-\left(
3/16\right) b$. Note that the right-hand sides of the $2$ ODEs (\ref{4ii13d}%
) are \textit{polynomial} only if $a=0.$

\bigskip

\subsection{Model 4.(i)1.4a}

\begin{subequations}
\label{4i14a}
\begin{eqnarray}
&\dot{x}_{1}=x_{1}\left\{ a+\left( b/3\right) \cdot \right. &  \notag \\
&\left. \cdot \left[ 243\left( x_{1}\right) ^{4}+648\left( x_{1}\right)
^{3}x_{2}-106\left( x_{1}x_{2}\right) ^{2}-16x_{1}\left( x_{2}\right)
^{3}-\left( x_{2}\right) ^{4}\right] \right\} ~,&
\end{eqnarray}%
\begin{eqnarray}
&\dot{x}_{2}=x_{2}\left\{ a-b\cdot \right. &  \notag \\
&\left. \cdot \left[ 243\left( x_{1}\right) ^{4}-376\left( x_{1}\right)
^{3}x_{2}-106\left( x_{1}x_{2}\right) ^{2}-16x_{1}\left( x_{2}\right)
^{3}-\left( x_{2}\right) ^{4}\right] \right\} ~;&
\end{eqnarray}%
$x_{1}\left( t\right) $ and $x_{2}\left( t\right) $ are related to $%
y_{1}\left( t\right) $ and $y_{4}\left( t\right) $ by (\ref{M4N2mu31ym});
and the variables $y_{1}\left( t\right) $ and $y_{4}\left( t\right) $ evolve
according to (\ref{A1SolvSyst}), the \textit{explicit} solution of which is
given by the relevant formulas in \textbf{Subsection Case A.1 }of \textbf{%
Appendix A} with $\tilde{m}_{1}=1,$ $\tilde{m}_{2}=4,$ $L=1,$ $\alpha
_{0}=a, $ $\alpha _{1}=b,$ $\beta _{0}=4a,$ $\beta _{1}=2^{10}b=1024b$.

\bigskip

\subsection{Model 4.(ii)1.4a}

\end{subequations}
\begin{eqnarray}
&&\dot{x}_{n}=x_{n}\left\{ a+b\cdot \right.  \notag \\
&&\left. \cdot \left[ \left( x_{n}\right) ^{4}+6\left( x_{n}\right)
^{3}x_{n+1}+16\left( x_{1}x_{2}\right) ^{2}-6x_{n}\left( x_{n+1}\right)
^{3}-\left( x_{n+1}\right) ^{4}\right] \right\} ~,  \notag \\
&&n=1,2~~\func{mod}\left[ 2\right] ~,
\end{eqnarray}%
$x_{1}\left( t\right) $ and $x_{2}\left( t\right) $ are related to $%
y_{1}\left( t\right) $ and $y_{4}\left( t\right) $ by (\ref{M4N2mu22ym});
and the variables $y_{1}\left( t\right) $ and $y_{4}\left( t\right) $ evolve
according to (\ref{A1SolvSyst}), the \textit{explicit} solution of which is
given by the relevant formulas in \textbf{Subsection Case A.1 }of \textbf{%
Appendix A} with $\tilde{m}_{1}=1,$ $\tilde{m}_{2}=4,$ $L=1,$ $\alpha
_{0}=a, $ $\alpha _{1}=b/16,$ $\beta _{0}=4a,$ $\beta _{1}=2^{6}b=64b$.

\bigskip

\subsection{Model 4.(i)1.4b}

\begin{subequations}
\label{4i14b}
\begin{eqnarray}
&&\dot{x}_{1}=x_{1}\left( a_{0}+a_{1}X+a_{2}X^{2}\right)  \notag \\
&&-\left[ \frac{37\left( x_{1}\right) ^{2}+10x_{1}x_{2}+\left( x_{2}\right)
^{2}}{3\left( x_{1}\right) ^{2}}\right] \left(
b_{0}+b_{1}X+b_{2}X^{2}+b_{3}X^{3}\right) ~,
\end{eqnarray}%
\begin{eqnarray}
&&\dot{x}_{2}=x_{2}\left( a_{0}+a_{1}X+a_{2}X^{2}\right)  \notag \\
&&-\left[ \frac{27\left( x_{1}\right) ^{2}-10x_{1}x_{2}-\left( x_{2}\right)
^{2}}{\left( x_{1}\right) ^{2}}\right] \left(
b_{0}+b_{1}X+b_{2}X^{2}+b_{3}X^{3}\right) ~;
\end{eqnarray}%
\begin{equation}
X\equiv 3x_{1}+x_{2}~;
\end{equation}%
$x_{1}\left( t\right) $ and $x_{2}\left( t\right) $ are related to $%
y_{1}\left( t\right) $ and $y_{4}\left( t\right) $ by (\ref{M4N2mu31ym});
and the variables $y_{1}\left( t\right) $ and $y_{4}\left( t\right) $ evolve
according to (\ref{A2SolvSyst}), the \textit{explicit} solution of which is
given by the relevant formulas in \textbf{Subsection} \textbf{Case A.2} of
\textbf{Appendix A} with $\tilde{m}_{1}=1,$ $\tilde{m}_{2}=4,$ $L=3,$ $%
\alpha _{0}=64b_{0};$ $\alpha _{\ell }=\left( -1\right) ^{\ell -1}\left(
a_{\ell -1}-64b_{\ell }\right) ,$ $\beta _{\ell }=\left( -1\right) ^{\ell
-1}4a_{\ell -1},$ $\ell =1,2,3;$ $\gamma _{\ell }=\left( -1\right) ^{\ell
}b_{\ell },$ $\ell =0,1,2,3.$ Note that the right-hand sides of these $2$
ODEs, (\ref{4i14b}), are both \textit{polynomial} only if all the $4$
parameters $b_{\ell }$ vanish, $b_{\ell }=0,$ $\ell =0,1,2,3$.

\bigskip

\subsection{Model 4.(ii)1.4b}

\end{subequations}
\begin{eqnarray}
\dot{x}_{n}=x_{n}\left( a_{0}+a_{1}X+a_{2}X^{2}\right) +\left[ \frac{\left(
x_{n}\right) ^{2}-4x_{1}x_{2}-\left( x_{n+1}\right) ^{2}}{x_{1}x_{2}}\right]
\cdot &&  \notag \\
\cdot \left( b_{0}+b_{1}X+b_{2}X^{2}+b_{3}X^{3}\right) ~,~~~X\equiv
x_{1}+x_{2}~,~~~n=1,2~~\func{mod}\left[ 2\right] ~; &&  \label{4i14bb}
\end{eqnarray}%
$x_{1}\left( t\right) $ and $x_{2}\left( t\right) $ are related to $%
y_{1}\left( t\right) $ and $y_{4}\left( t\right) $ by (\ref{M4N2mu22ym});
and the variables $y_{1}\left( t\right) $ and $y_{4}\left( t\right) $ evolve
according to (\ref{A2SolvSyst}), the \textit{explicit} solution of which is
given by the relevant formulas in \textbf{Subsection} \textbf{Case A.2} of
\textbf{Appendix A} with $\tilde{m}_{1}=1,$ $\tilde{m}_{2}=4,$ $L=3,$ $%
\alpha _{0}=16b_{0};$ $\alpha _{\ell }=\left( -2\right) ^{1-\ell }a_{\ell
-1}+\left( -2\right) ^{4-\ell }b_{\ell },$ $\beta _{\ell }=\left( -2\right)
^{3-\ell }a_{\ell -1},$ $\ell =1,2,3;$ $\gamma _{\ell }=\left( -2\right)
^{-2-\ell }b_{\ell },$ $\ell =0,1,2,3.$ Note that the right-hand sides of
these $2$ ODEs, (\ref{4i14bb}), are both \textit{polynomial} only if all the
$4$ parameters $b_{\ell }$ vanish, $b_{\ell }=0,$ $\ell =1,2,3$.

\bigskip

\subsection{Model 4.(i)1.4c{}}

\begin{equation}
\dot{x}_{n}=x_{n}\left( a+bX+cX^{2}\right) ~,~~~X\equiv \left( x_{1}\right)
^{3}x_{2}~,~~~n=1,2~;  \label{4i14c}
\end{equation}%
$x_{1}\left( t\right) $ and $x_{2}\left( t\right) $ are related to $%
y_{1}\left( t\right) $ and $y_{4}\left( t\right) $ by (\ref{M4N2mu31ym});
and the variables $y_{1}\left( t\right) $ and $y_{4}\left( t\right) $ evolve
according to (\ref{A2SolvSyst}), the \textit{explicit} solution of which is
given by the relevant formulas in \textbf{Subsection Case A.2} of \textbf{%
Appendix A} with $\tilde{m}_{1}=4,$ $\tilde{m}_{2}=1,$ $L=3,$ $\alpha
_{0}=0, $ $\alpha _{1}=4a,$ $\alpha _{2}=4b,$ $\alpha _{3}=4c,$ $\beta
_{1}=a,$ $\beta _{2}=b,$ $\beta _{3}=c,$ $\gamma _{\ell }=0.$

\bigskip

\subsection{Model 4.(ii)1.4c{}}

\begin{equation}
\dot{x}_{n}=x_{n}\left( a+bX+cX^{2}\right) ~,~X\equiv \left(
x_{1}x_{2}\right) ^{2}~,~~~n=1,2~;  \label{4ii14c}
\end{equation}%
$x_{1}\left( t\right) $ and $x_{2}\left( t\right) $ are related to $%
y_{1}\left( t\right) $ and $y_{4}\left( t\right) $ by (\ref{M4N2mu22ym});
and the variables $y_{1}\left( t\right) $ and $y_{4}\left( t\right) $ evolve
according to (\ref{A2SolvSyst}), the \textit{explicit} solution of which is
given by the relevant formulas in \textbf{Subsection Case A.2} of \textbf{%
Appendix A} with $\tilde{m}_{1}=4,$ $\tilde{m}_{2}=1,$ $L=3,$ $\alpha
_{0}=0, $ $\alpha _{1}=4a,$ $\alpha _{2}=4b,$ $\alpha _{3}=4c,$ $\beta
_{1}=a,$ $\beta _{2}=b,$ $\beta _{3}=c,$ $\gamma _{\ell }=0.$

\bigskip

\subsection{Model 4.(i)1.4d}

\begin{subequations}
\label{4i14d}
\begin{eqnarray}
&&\dot{x}_{1}=\left[ 3\left( x_{1}\right) ^{2}\right] ^{-1}\left\{ a\left[
37\left( x_{1}\right) ^{2}+10x_{1}x_{2}+\left( x_{2}\right) ^{2}\right]
\right.  \notag \\
&&+b\left[ 2187\left( x_{1}\right) ^{6}-9094\left( x_{1}\right)
^{5}x_{2}-3991\left( x_{1}\right) ^{4}\left( x_{2}\right) ^{2}-1156\left(
x_{1}x_{2}\right) ^{3}\right.  \notag \\
&&\left. \left. -211\left( x_{1}\right) ^{2}\left( x_{2}\right)
^{4}-22x_{1}\left( x_{2}\right) ^{5}-\left( x_{2}\right) ^{6}\right]
\right\} ~,
\end{eqnarray}%
\begin{eqnarray}
&&\dot{x}_{2}=\left[ \left( x_{1}\right) ^{2}\right] ^{-1}\left\{ a\left[
27\left( x_{1}\right) ^{2}-10x_{1}x_{2}-\left( x_{2}\right) ^{2}\right]
\right.  \notag \\
&&-b\left[ 2187\left( x_{1}\right) ^{6}+7290\left( x_{1}\right)
^{5}x_{2}-3991\left( x_{1}\right) ^{4}\left( x_{2}\right) ^{2}-1156\left(
x_{1}x_{2}\right) ^{3}\right.  \notag \\
&&\left. \left. -211\left( x_{1}\right) ^{2}\left( x_{2}\right)
^{4}-22x_{1}\left( x_{2}\right) ^{5}-\left( x_{2}\right) ^{6}\right]
\right\} ~;
\end{eqnarray}%
$x_{1}\left( t\right) $ and $x_{2}\left( t\right) $ are related to $%
y_{1}\left( t\right) $ and $y_{4}\left( t\right) $ by (\ref{M4N2mu31ym});
and the variables $y_{1}\left( t\right) $ and $y_{4}\left( t\right) $ evolve
according to (\ref{A3SolvSysta}), the \textit{explicit} solution of which is
given by the relevant formulas in \textbf{Subsection} \textbf{Case A.3.3} of
\textbf{Appendix A} with $\tilde{m}_{1}=1,$ $\tilde{m}_{2}=4,$ $\alpha
_{0}=-2^{6}a=-64a;$ $\alpha _{1}=2^{14}b=1638b,$ $\beta _{0}=-a,$ $\beta
_{1}=b.$ Note that the right-hand sides of these $2$ ODEs, (\ref{4i14d}),
are \textit{not} \textit{polynomial}.

\bigskip

\subsection{Model 4.(ii)1.4d}

\end{subequations}
\begin{eqnarray}
&&\dot{x}_{n}=\left( x_{1}x_{2}\right) ^{-1}\left\{ a\left[ \left(
x_{n}\right) ^{2}-4x_{1}x_{2}-\left( x_{n+1}\right) ^{2}\right] \right.
\notag \\
&&+b\left[ \left( x_{n}\right) ^{6}+8\left( x_{n}\right)
^{5}x_{n+1}+29\left( x_{n}\right) ^{4}\left( x_{n+1}\right) ^{2}-64\left(
x_{1}x_{2}\right) ^{3}\right.  \notag \\
&&\left. \left. -29\left( x_{n}\right) ^{2}\left( x_{n+1}\right)
^{4}-8x_{n}\left( x_{n+1}\right) ^{5}-\left( x_{n+1}\right) ^{6}\right]
\right\} ~,  \notag \\
&&n=1,2~~\func{mod}\left[ 2\right] ~;  \label{4ii14d}
\end{eqnarray}%
$x_{1}\left( t\right) $ and $x_{2}\left( t\right) $ are related to $%
y_{1}\left( t\right) $ and $y_{4}\left( t\right) $ by (\ref{M4N2mu22ym});
and the variables $y_{1}\left( t\right) $ and $y_{4}\left( t\right) $ evolve
according to (\ref{A3SolvSysta}), the \textit{explicit} solution of which is
given by the relevant formulas in \textbf{Subsection} \textbf{Case A.3.3} of
\textbf{Appendix A} with $\tilde{m}_{1}=1,$ $\tilde{m}_{2}=4,$ $\alpha
_{0}=2^{4}a=16a;$ $\alpha _{1}=2^{8}b=256b,$ $\beta _{0}=2^{-2}a=a/4,$ $%
\beta _{1}=2^{-6}b=b/64.$ Note that the right-hand sides of these $2$ ODEs, (%
\ref{4i14d}), are \textit{not} \textit{polynomial}.

\bigskip

\subsection{Model 4.(i)2.3a}

\begin{subequations}
\begin{equation}
\dot{x}_{1}=x_{1}\left\{ a+b\left( x_{1}\right) ^{3}\left[ 3\left(
x_{1}\right) ^{3}+10\left( x_{1}\right) ^{2}x_{2}+7x_{1}\left( x_{2}\right)
^{2}-4\left( x_{2}\right) ^{3}\right] \right\} ~,
\end{equation}%
\begin{equation}
\dot{x}_{2}=ax_{2}-b\left( x_{1}\right) ^{3}\left[ 2\left( x_{1}\right)
^{4}+7\left( x_{1}\right) ^{3}x_{2}-17x_{1}\left( x_{2}\right) ^{3}-8\left(
x_{2}\right) ^{4}\right] ~;
\end{equation}%
$x_{1}\left( t\right) $ and $x_{2}\left( t\right) $ are related to $%
y_{2}\left( t\right) $ and $y_{3}\left( t\right) $ by (\ref{M4N2mu31ym});
and the variables $y_{2}\left( t\right) $ and $y_{3}\left( t\right) $ evolve
according to (\ref{A1SolvSyst}), the \textit{explicit} solution of which is
given by the relevant formulas in \textbf{Subsection} \textbf{Case A.1} of
\textbf{Appendix A} with $\tilde{m}_{1}=2,$ $\tilde{m}_{2}=3,$ $L=1,$ $%
\alpha _{0}=2a,$ $\alpha _{1}=\left( 4/27\right) b,$ $\beta _{0}=3a,$ $\beta
_{1}=3b.$

\bigskip

\subsection{Model 4.(ii)2.3a}

\end{subequations}
\begin{eqnarray}
&&\dot{x}_{n}=x_{n}\left\{ a+b\left[ \left( x_{1}\right)
^{2}+x_{1}x_{2}+\left( x_{2}\right) ^{2}\right] ^{-1}\cdot \right.  \notag \\
&&\cdot \left[ \left( x_{n}\right) ^{8}+19\left( x_{n}\right)
^{7}x_{n+1}+151\left( x_{n}\right) ^{6}\left( x_{n+1}\right) ^{2}+331\left(
x_{n}\right) ^{5}\left( x_{n+1}\right) ^{3}\right.  \notag \\
&&+259\left( x_{1}x_{2}\right) ^{4}+13\left( x_{n}\right) ^{3}\left(
x_{n+1}\right) ^{5}-89\left( x_{n}\right) ^{2}\left( x_{n+1}\right)
^{6}-35x_{n}\left( x_{n+1}\right) ^{7}  \notag \\
&&\left. \left. -2\left( x_{n+1}\right) ^{8}\right] \right\} ~,~~~n=1,2~~%
\func{mod}\left[ 2\right] ~;
\end{eqnarray}%
$x_{1}\left( t\right) $ and $x_{2}\left( t\right) $ are related to $%
y_{2}\left( t\right) $ and $y_{3}\left( t\right) $ by (\ref{M4N2mu22ym});
and the variables $y_{2}\left( t\right) $ and $y_{3}\left( t\right) $ evolve
according to (\ref{A1SolvSyst}), the \textit{explicit} solution of which is
given by the relevant formulas in \textbf{Subsection} \textbf{Case A.1} of
\textbf{Appendix A} with $\tilde{m}_{1}=2,$ $\tilde{m}_{2}=3,$ $L=1,$ $%
\alpha _{0}=2a,$ $\alpha _{1}=2b,$ $\beta _{0}=3a,$ $\beta _{1}=\left(
81/2\right) b.$ Note that the right-hand sides of these ODEs is \textit{not}
polynomial.

\bigskip

\subsection{Model 4.(i)2.3b}

\begin{equation}
\dot{x}_{n}=x_{n}\left( a+bX+cX^{2}\right) ~,~~~X\equiv x_{1}\left(
x_{1}+x_{2}\right) ~,~~~n=1,2~;  \label{4i23b}
\end{equation}%
$x_{1}\left( t\right) $ and $x_{2}\left( t\right) $ are related to $%
y_{2}\left( t\right) $ and $y_{3}\left( t\right) $ by (\ref{M4N2mu31ym});
and the variables $y_{2}\left( t\right) $ and $y_{3}\left( t\right) $ evolve
according to (\ref{A2SolvSyst}), the \textit{explicit} solution of which is
given by the relevant formulas in \textbf{Subsection} \textbf{Case A.2} of
\textbf{Appendix A} with $\tilde{m}_{1}=2,$ $\tilde{m}_{2}=3,$ $L=3,$ $%
\alpha _{0}=0~,$ $\alpha _{1}=2a,$ $\alpha _{2}=\left( 2/3\right) b,$ $%
\alpha _{3}=\left( 2/9\right) c,$ $\beta _{1}=3a,$ $\beta _{2}=b,~\beta
_{3}=c/3,~\gamma _{\ell }=0.$

\bigskip

\subsection{Model 4.(ii)2.3b}

\begin{equation}
\dot{x}_{n}=x_{n}\left( a+bX+cX^{2}\right) ~,~~~X\equiv \left( x_{1}\right)
^{2}+4x_{1}x_{2}+\left( x_{2}\right) ^{2}~,~~~n=1,2~;  \label{4ii23b}
\end{equation}%
$x_{1}\left( t\right) $ and $x_{2}\left( t\right) $ are related to $%
y_{2}\left( t\right) $ and $y_{3}\left( t\right) $ by (\ref{M4N2mu22ym});
and the variables $y_{2}\left( t\right) $ and $y_{3}\left( t\right) $ evolve
according to (\ref{A2SolvSyst}), the \textit{explicit} solution of which is
given by the relevant formulas in \textbf{Subsection} \textbf{Case A.2} of
\textbf{Appendix A} with $\tilde{m}_{1}=2,$ $\tilde{m}_{2}=3,$ $L=3,$ $%
\alpha _{0}=0~,$ $\alpha _{1}=2a,$ $\alpha _{2}=2b,$ $\alpha _{3}=2c,$ $%
\beta _{1}=3a,$ $\beta _{2}=3b,~\beta _{3}=3c,~\gamma _{\ell }=0.$

\bigskip

\subsection{Model 4.(i)2.3c}

\begin{equation}
\dot{x}_{n}=x_{n}\left( a+bX+cX^{2}\right) ~,~~~X\equiv \left( x_{1}\right)
^{2}\left( x_{1}+3x_{2}\right) ~,~~~n=1,2~;  \label{4i23cc}
\end{equation}%
$x_{1}\left( t\right) $ and $x_{2}\left( t\right) $ are related to $%
y_{2}\left( t\right) $ and $y_{3}\left( t\right) $ by (\ref{M4N2mu31ym});
and the variables $y_{2}\left( t\right) $ and $y_{3}\left( t\right) $ evolve
according to (\ref{A2SolvSyst}), the \textit{explicit} solution of which is
given by the relevant formulas in \textbf{Subsection} \textbf{Case A.2} of
\textbf{Appendix A} with $\tilde{m}_{1}=3,$ $\tilde{m}_{2}=2,$ $L=3,$ $%
\alpha _{0}=0~,$ $\alpha _{1}=3a,$ $\alpha _{2}=-3b,$ $\alpha _{3}=3c,$ $%
\beta _{1}=2a,$ $\beta _{2}=-2b,$ $\beta _{3}=3c,$ $\gamma _{\ell }=0.$

\bigskip

\subsection{Model 4.(ii)2.3c}

\begin{equation}
\dot{x}_{n}=x_{n}\left( a+bX+cX^{2}\right) ~,~~~X\equiv x_{1}x_{2}\left(
x_{1}+x_{2}\right) ~,~~~n=1,2~;  \label{4ii23c}
\end{equation}%
$x_{1}\left( t\right) $ and $x_{2}\left( t\right) $ are related to $%
y_{2}\left( t\right) $ and $y_{3}\left( t\right) $ by (\ref{M4N2mu22ym});
and the variables $y_{2}\left( t\right) $ and $y_{3}\left( t\right) $ evolve
according to (\ref{A2SolvSyst}), the \textit{explicit} solution of which is
given by the relevant formulas in \textbf{Subsection} \textbf{Case A.2} of
\textbf{Appendix A} with $\tilde{m}_{1}=3,$ $\tilde{m}_{2}=2,$ $L=3,$ $%
\alpha _{0}=0~,$ $\alpha _{1}=3a,$ $\alpha _{2}=-\left( 3/2\right) b,$ $%
\alpha _{3}=\left( 3/4\right) c,$ $\beta _{1}=2a,$ $\beta _{2}=-b,$ $\beta
_{3}=c/2,$ $\gamma _{\ell }=0.$

\bigskip

\subsection{Model 4.(i)2.4a}

\begin{subequations}
\begin{eqnarray}
&&\dot{x}_{1}=x_{1}\left\{ a+b\left( x_{1}\right) ^{2}\left[ \left(
x_{1}\right) ^{2}+4x_{1}x_{2}-\left( x_{2}\right) ^{2}\right] \right.  \notag
\\
&&\left. +c\left( x_{1}\right) ^{4}\left[ \left( x_{1}\right) ^{4}+6\left(
x_{1}\right) ^{3}x_{2}+16\left( x_{1}x_{2}\right) ^{2}-6x_{1}\left(
x_{2}\right) ^{3}-\left( x_{2}\right) ^{4}\right] \right\} ~,
\end{eqnarray}%
\begin{eqnarray}
&&\dot{x}_{2}=x_{2}\left\{ a-b\left( x_{1}\right) ^{2}\left[ 3\left(
x_{1}\right) ^{2}-4x_{1}x_{2}-3\left( x_{2}\right) ^{2}\right] \right.
\notag \\
&&\left. +c\left( x_{1}\right) ^{4}\left[ -3\left( x_{1}\right)
^{4}-18\left( x_{1}\right) ^{3}x_{2}+16\left( x_{1}x_{2}\right)
^{2}+18x_{1}\left( x_{2}\right) ^{3}+3\left( x_{2}\right) ^{4}\right]
\right\} ~,  \notag \\
&&
\end{eqnarray}%
$x_{1}\left( t\right) $ and $x_{2}\left( t\right) $ are related to $%
y_{2}\left( t\right) $ and $y_{4}\left( t\right) $ by (\ref{M4N2mu31ym});
and the variables $y_{2}\left( t\right) $ and $y_{4}\left( t\right) $ evolve
according to (\ref{A1SolvSyst}), the \textit{explicit} solution of which is
given by the relevant formulas in \textbf{Subsection} \textbf{Case A.1} of
\textbf{Appendix A} with $\tilde{m}_{1}=2,$ $\tilde{m}_{2}=4,$ $L=2,$ $%
\alpha _{0}=2a,$ $\alpha _{1}=\left( 2/9\right) b,$ $\alpha _{2}=\left(
2/81\right) c,$ $\beta _{0}=4a,$ $\beta _{1}=16b,$ $\beta _{2}=64c.$

\bigskip

\subsection{Model 4.(ii)2.4a}

\end{subequations}
\begin{eqnarray}
&&\dot{x}_{n}=x_{n}\left\{ a+\left( x_{1}+x_{2}\right) ^{-1}\cdot \left\{ b%
\left[ \left( x_{n}\right) ^{5}+13\left( x_{n}\right) ^{4}x_{n+1}+64\left(
x_{n}\right) ^{3}\left( x_{n+1}\right) ^{2}\right. \right. \right.  \notag \\
&&\left. +8\left( x_{n}\right) ^{2}\left( x_{n+1}\right) ^{3}-13x_{n}\left(
x_{n+1}\right) ^{4}-\left( x_{n+1}\right) ^{5}\right]  \notag \\
&&+c\left[ \left( x_{n}\right) ^{9}+21\left( x_{n}\right)
^{8}x_{n+1}+186\left( x_{n}\right) ^{7}\left( x_{n+1}\right) ^{2}+906\left(
x_{n}\right) ^{6}\left( x_{n+1}\right) ^{3}\right.  \notag \\
&&+2676\left( x_{n}\right) ^{5}\left( x_{n+1}\right) ^{4}-84\left(
x_{n}\right) ^{4}\left( x_{n+1}\right) ^{5}-906\left( x_{n}\right)
^{3}\left( x_{n+1}\right) ^{6}  \notag \\
&&\left. \left. \left. -186\left( x_{n}\right) ^{2}\left( x_{n+1}\right)
^{7}-21x_{n}\left( x_{n+1}\right) ^{8}-\left( x_{n+1}\right) ^{9}\right]
\right\} \right\} ~,  \notag \\
&&n=1,2~~\func{mod}\left[ 2\right] ~;  \label{4ii24a}
\end{eqnarray}%
$x_{1}\left( t\right) $ and $x_{2}\left( t\right) $ are related to $%
y_{2}\left( t\right) $ and $y_{4}\left( t\right) $ by (\ref{M4N2mu22ym});
and the variables $y_{2}\left( t\right) $ and $y_{4}\left( t\right) $ evolve
according to (\ref{A1SolvSyst}), the \textit{explicit} solution of which is
given by the relevant formulas in \textbf{Subsection} \textbf{Case A.1} of
\textbf{Appendix A} with $\tilde{m}_{1}=2,$ $\tilde{m}_{2}=4,$ $L=2,$ $%
\alpha _{0}=2a,$ $\alpha _{1}=2b,$ $\alpha _{2}=-2c,$ $\beta _{0}=4a,$ $%
\beta _{1}=144b,$ $\beta _{2}=-2^{6}3^{4}c=-5184c.$ Note that the right-hand
sides of these ODEs are \textit{not} polynomial, except for the trivial case
with $b=c=0.$

\bigskip

\subsection{Model 4.(i)2.4b}

\begin{subequations}
\label{4i24b}
\begin{equation}
\dot{x}_{1}=x_{1}\left( a_{0}+a_{1}X+a_{2}X^{2}\right) +\left( x_{1}\right)
^{-1}\left( b_{0}+b_{1}X+b_{2}X^{2}+b_{3}X^{3}\right) ~,
\end{equation}%
\begin{equation}
\dot{x}_{2}=x_{2}\left( a_{0}+a_{1}X+a_{2}X^{2}\right) +\frac{\left(
2x_{1}-x_{2}\right) }{\left( x_{1}\right) ^{2}}\left(
b_{0}+b_{1}X+b_{2}X^{2}+b_{3}X^{3}\right) ~,
\end{equation}%
\begin{equation}
X\equiv x_{1}\left( x_{1}+x_{2}\right) ~;
\end{equation}%
$x_{1}\left( t\right) $ and $x_{2}\left( t\right) $ are related to $%
y_{2}\left( t\right) $ and $y_{4}\left( t\right) $ by (\ref{M4N2mu31ym});
and the variables $y_{2}\left( t\right) $ and $y_{4}\left( t\right) $ evolve
according to (\ref{A2SolvSyst}), the \textit{explicit} solution of which is
given by the relevant formulas in \textbf{Subsection} \textbf{Case A.2} of
\textbf{Appendix A} with $\tilde{m}_{1}=2,$ $\tilde{m}_{2}=4,$ $L=3,$ $%
\alpha _{0}=12b_{0};$ $\alpha _{\ell }=2\left( 3^{1-\ell }\right) a_{\ell
-1}+4\left( 3^{1-\ell }\right) b_{\ell },$ $\beta _{\ell }=4\left( 3^{1-\ell
}\right) a_{\ell -1},$ $\ell =1,2,3;$ $\gamma _{\ell }=2\left( 3^{-1-\ell
}\right) b_{\ell },$ $\ell =0,1,2,3.$ Note that the right-hand sides of
these $2$ ODEs, (\ref{4i14b}), are both \textit{polynomial} only if \textit{%
all} the $4$ parameters $b_{\ell }$ vanish, $b_{\ell }=0,$ $\ell =0,1,2,3$.

\bigskip

\subsection{Model 4.(ii)2.4b}

\end{subequations}
\begin{eqnarray}
&&\dot{x}_{n}=x_{n}\left\{ a_{0}+a_{1}X+a_{2}X^{2}\right.  \notag \\
&&\left. +\left[ \frac{-2\left( x_{n}\right) ^{2}+7x_{n}x_{n+1}+\left(
x_{n+1}\right) ^{2}}{x_{1}x_{2}\left( x_{1}+x_{2}\right) }\right] \left(
b_{0}+b_{1}X+b_{2}X^{2}+b_{3}X^{3}\right) \right\} ~,  \notag \\
&&X\equiv \left( x_{1}\right) ^{2}+4x_{1}x_{2}+\left( x_{2}\right) ^{2}~;
\label{4ii24b}
\end{eqnarray}%
$x_{1}\left( t\right) $ and $x_{2}\left( t\right) $ are related to $%
y_{2}\left( t\right) $ and $y_{4}\left( t\right) $ by (\ref{M4N2mu22ym});
and the variables $y_{2}\left( t\right) $ and $y_{4}\left( t\right) $ evolve
according to (\ref{A2SolvSyst}), the \textit{explicit} solution of which is
given by the relevant formulas in \textbf{Subsection} \textbf{Case A.2} of
\textbf{Appendix A} with $\tilde{m}_{1}=2,$ $\tilde{m}_{2}=4,$ $L=3,$ $%
\alpha _{0}=36b_{0};$ $\alpha _{\ell }=2a_{\ell -1}+36b_{\ell },$ $\beta
_{\ell }=4a_{\ell -1},$ $\ell =1,2,3;$ $\gamma _{\ell }=2b_{\ell },$ $\ell
=0,1,2,3.$ Note that the right-hand sides of these $2$ ODEs, (\ref{4ii24b}),
are both \textit{polynomial} only if \textit{all} the $4$ parameters $%
b_{\ell }$ vanish, $b_{\ell }=0,$ $\ell =0,1,2,3$.

\bigskip

\subsection{Model 4.(i)2.4c}

\begin{equation}
\dot{x}_{n}=x_{n}\left( a+bX+cX^{2}\right) ~,~~~X\equiv \left( x_{1}\right)
^{3}x_{2}~,~~~n=1,2~;  \label{4i24c}
\end{equation}%
$x_{1}\left( t\right) $ and $x_{2}\left( t\right) $ are related to $%
y_{2}\left( t\right) $ and $y_{4}\left( t\right) $ by (\ref{M4N2mu31ym});
and the variables $y_{2}\left( t\right) $ and $y_{4}\left( t\right) $ evolve
according to (\ref{A2SolvSyst}), the \textit{explicit} solution of which is
given by the relevant formulas in \textbf{Subsection} \textbf{Case A.2} of
\textbf{Appendix A} with $\tilde{m}_{1}=4,$ $\tilde{m}_{2}=2,$ $L=3,$ $%
\alpha _{0}=0~,$ $\alpha _{1}=4a,$ $\alpha _{2}=4b,$ $\alpha _{3}=4c,$ $%
\beta _{1}=2a,$ $\beta _{2}=2b,~\beta _{3}=2c,~\gamma _{\ell }=0.$

\bigskip

\subsection{Model 4.(ii)2.4c}

\begin{equation}
\dot{x}_{n}=x_{n}\left( a+bX+cX^{2}\right) ~,~~~X\equiv \left(
x_{1}x_{2}\right) ^{2}~,~~~n=1,2~;  \label{4ii24c}
\end{equation}%
$x_{1}\left( t\right) $ and $x_{2}\left( t\right) $ are related to $%
y_{2}\left( t\right) $ and $y_{4}\left( t\right) $ by (\ref{M4N2mu22ym});
and the variables $y_{2}\left( t\right) $ and $y_{4}\left( t\right) $ evolve
according to (\ref{A2SolvSyst}), the \textit{explicit} solution of which is
given by the relevant formulas in \textbf{Subsection} \textbf{Case A.2} of
\textbf{Appendix A} with $\tilde{m}_{1}=4,$ $\tilde{m}_{2}=2,$ $L=3,$ $%
\alpha _{0}=0~,$ $\alpha _{1}=4a,$ $\alpha _{2}=4b,$ $\alpha _{3}=4c,$ $%
\beta _{1}=2a,$ $\beta _{2}=2b,~\beta _{3}=2c,~\gamma _{\ell }=0.$

\bigskip

\subsection{Model 4.(i)2.4d}

\begin{subequations}
\label{4i24d}
\begin{equation}
\dot{x}_{1}=\left( x_{1}\right) ^{-1}\left\{ a+b\left( x_{1}\right) ^{2}%
\left[ \left( x_{1}\right) ^{2}-4x_{1}x_{2}-\left( x_{2}\right) ^{2}\right]
\right\} ~,
\end{equation}%
\begin{equation}
\dot{x}_{2}=\left( x_{1}\right) ^{-2}\left\{ a\left( 2x_{1}-x_{2}\right)
-b\left( x_{1}\right) ^{2}\left[ 2\left( x_{1}\right) ^{3}+9\left(
x_{1}\right) ^{2}x_{2}-6x_{1}\left( x_{2}\right) ^{2}-\left( x_{2}\right)
^{3}\right] \right\} ~,
\end{equation}%
$x_{1}\left( t\right) $ and $x_{2}\left( t\right) $ are related to $%
y_{2}\left( t\right) $ and $y_{4}\left( t\right) $ by (\ref{M4N2mu31ym});
and the variables $y_{2}\left( t\right) $ and $y_{4}\left( t\right) $ evolve
according to (\ref{A3SolvSysta}), the \textit{explicit} solution of which is
given by the relevant formulas in \textbf{Subsection} \textbf{Case A.3.1} of
\textbf{Appendix A} with $\tilde{m}_{1}=2,$ $\tilde{m}_{2}=4,$ $\alpha
_{0}=12a;$ $\alpha _{1}=-48b,$ $\beta _{0}=\left( 2/3\right) a,$ $\beta
_{1}=-\left( 2/27\right) b.$ Note that the right-hand sides of these $2$
ODEs, (\ref{4i14d}), are \textit{not} \textit{polynomial}, unless $a$
vanishes.

\bigskip

\subsection{Model 4.(ii)2.4d}

\end{subequations}
\begin{eqnarray}
&&\dot{x}_{n}=\left[ x_{1}x_{2}\left( x_{1}+x_{2}\right) \right]
^{-1}\left\{ a\left[ 2\left( x_{n}\right) ^{2}-7x_{1}x_{2}-\left(
x_{n+1}\right) ^{2}\right] \right.  \notag \\
&&+b\left[ 2\left( x_{n}\right) ^{6}+27\left( x_{n}\right)
^{5}x_{n+1}+141\left( x_{n}\right) ^{4}\left( x_{n+1}\right) ^{2}-280\left(
x_{1}x_{2}\right) ^{3}\right.  \notag \\
&&\left. \left. -90\left( x_{n}\right) ^{2}\left( x_{n+1}\right)
^{4}-15x_{n}\left( x_{n+1}\right) ^{5}-\left( x_{n+1}\right) ^{6}\right]
\right\} ~,  \notag \\
&&n=1,2~~\func{mod}\left[ 2\right] ~;  \label{4ii24d}
\end{eqnarray}%
$x_{1}\left( t\right) $ and $x_{2}\left( t\right) $ are related to $%
y_{2}\left( t\right) $ and $y_{4}\left( t\right) $ by (\ref{M4N2mu22ym});
and the variables $y_{2}\left( t\right) $ and $y_{4}\left( t\right) $ evolve
according to (\ref{A3SolvSysta}), the \textit{explicit} solution of which is
given by the relevant formulas in \textbf{Subsection} \textbf{Case A.3.1} of
\textbf{Appendix A} with $\tilde{m}_{1}=2,$ $\tilde{m}_{2}=4,$ $\alpha
_{0}=-36a;$ $\alpha _{1}=-1296b,$ $\beta _{0}=-2a,$ $\beta _{1}=-9b.$ Note
that the right-hand sides of these $2$ ODEs, (\ref{4ii24d}), are \textit{not}
polynomial.

\bigskip

\subsection{Model 4.(i)3.4a}

\begin{subequations}
\label{4i34a}
\begin{eqnarray}
\dot{x}_{1}=x_{1}\left\{ a+b\left( x_{1}\right) ^{8}\cdot \right. &&  \notag
\\
\left. \cdot \left[ \left( x_{1}\right) ^{4}+16\left( x_{1}\right)
^{3}x_{2}+106\left( x_{1}x_{2}\right) ^{2}+376x_{1}\left( x_{2}\right)
^{3}-243\left( x_{2}\right) ^{4}\right] \right\} ~, &&
\end{eqnarray}%
\begin{eqnarray}
\dot{x}_{2}=x_{2}\left\{ a-b\left( x_{1}\right) ^{8}\cdot \right. &&  \notag
\\
\left. \cdot \left[ 3\left( x_{1}\right) ^{4}+48\left( x_{1}\right)
^{3}x_{2}+318\left( x_{1}x_{2}\right) ^{2}+104x_{1}\left( x_{2}\right)
^{3}-729\left( x_{2}\right) ^{4}\right] \right\} ~; &&
\end{eqnarray}%
$x_{1}\left( t\right) $ and $x_{2}\left( t\right) $ are related to $%
y_{3}\left( t\right) $ and $y_{4}\left( t\right) $ by (\ref{M4N2mu31ym});
and the variables $y_{3}\left( t\right) $ and $y_{4}\left( t\right) $ evolve
according to (\ref{A1SolvSyst}), the \textit{explicit} solution of which is
given by the relevant formulas in \textbf{Subsection} \textbf{Case A.1} of
\textbf{Appendix A} with $\tilde{m}_{1}=3,$ $\tilde{m}_{2}=4,$ $L=1,$ $%
\alpha _{0}=3a,$ $\alpha _{1}=3b,$ $\beta _{0}=4a,$ $\beta
_{1}=2^{10}b=1024b.$

\bigskip

\subsection{Model 4.(ii)3.4a}

\end{subequations}
\begin{eqnarray}
\dot{x}_{n}=x_{n}\left\{ a+b\left( x_{1}x_{2}\right) ^{4}\cdot \right. &&
\notag \\
\left. \cdot \left[ 3\left( x_{n}\right) ^{4}+18\left( x_{n}\right)
^{3}x_{n+1}+16\left( x_{1}x_{2}\right) ^{2}-18x_{n}\left( x_{n+1}\right)
^{3}-3\left( x_{n+1}\right) ^{4}\right] \right\} ~, &&  \notag \\
n=1,2~~\func{mod}\left[ 2\right] ; &&  \label{4ii34c}
\end{eqnarray}%
$x_{1}\left( t\right) $ and $x_{2}\left( t\right) $ are related to $%
y_{3}\left( t\right) $ and $y_{4}\left( t\right) $ by (\ref{M4N2mu22ym});
and the variables $y_{3}\left( t\right) $ and $y_{4}\left( t\right) $ evolve
according to (\ref{A1SolvSyst}), the \textit{explicit} solution of which is
given by the relevant formulas in \textbf{Subsection} \textbf{Case A.1} of
\textbf{Appendix A} with $\tilde{m}_{1}=3,$ $\tilde{m}_{2}=4,$ $L=1,$ $%
\alpha _{0}=3a,$ $\alpha _{1}=\left( 3/16\right) b,$ $\beta _{0}=4a,$ $\beta
_{1}=64b.$

\bigskip

\subsection{Model 4.(i)3.4b}

\begin{equation}
\dot{x}_{n}=x_{n}\left( a+bX+cX^{2}\right) ~,~~~X\equiv \left( x_{1}\right)
^{3}x_{2}~,~~~n=1,2~;  \label{4i34b}
\end{equation}%
$x_{1}\left( t\right) $ and $x_{2}\left( t\right) $ are related to $%
y_{3}\left( t\right) $ and $y_{4}\left( t\right) $ by (\ref{M4N2mu31ym});
and the variables $y_{3}\left( t\right) $ and $y_{4}\left( t\right) $ evolve
according to (\ref{A2SolvSyst}), the \textit{explicit} solution of which is
given by the relevant formulas in \textbf{Subsection} \textbf{Case A.2} of
\textbf{Appendix A} with $\tilde{m}_{1}=4,$ $\tilde{m}_{2}=3,$ $L=3,$ $%
\alpha _{0}=0~,$ $\alpha _{1}=4a,$ $\alpha _{2}=4b,$ $\alpha _{3}=4c,$ $%
\beta _{1}=3a,$ $\beta _{2}=3b,~\beta _{3}=3c,~\gamma _{\ell }=0.$

\bigskip

\subsection{Model 4.(ii)3.4b}

\begin{equation}
\dot{x}_{n}=x_{n}\left[ a+bX+cX^{2}\right] ~,~~~X=\left( x_{1}x_{2}\right)
^{2}~~~n=1,2~;  \label{4ii34b}
\end{equation}%
$x_{1}\left( t\right) $ and $x_{2}\left( t\right) $ are related to $%
y_{3}\left( t\right) $ and $y_{4}\left( t\right) $ by (\ref{M4N2mu22ym});
and the variables $y_{3}\left( t\right) $ and $y_{4}\left( t\right) $ evolve
according to (\ref{A2SolvSyst}), the \textit{explicit} solution of which is
given by the relevant formulas in \textbf{Subsection} \textbf{Case A.2} of
\textbf{Appendix A} with $\tilde{m}_{1}=4,$ $\tilde{m}_{2}=3,$ $L=3,$ $%
\alpha _{0}=0~,$ $\alpha _{1}=4a,$ $\alpha _{2}=4b,$ $\alpha _{3}=4c,$ $%
\beta _{1}=3a,$ $\beta _{2}=3b,$ $\beta _{3}=3c,$ $\gamma _{\ell }=0.$

\bigskip

\subsection{Model 4.(i)3.4c}

\begin{equation}
\dot{x}_{n}=x_{n}\left[ a+bX+cX^{2}\right] ~,~~~X\equiv \left( x_{1}\right)
^{2}\left( x_{1}+3x_{2}\right) ~,~~~n=1,2~;  \label{4i34c}
\end{equation}%
$x_{1}\left( t\right) $ and $x_{2}\left( t\right) $ are related to $%
y_{3}\left( t\right) $ and $y_{4}\left( t\right) $ by (\ref{M4N2mu31ym});
and the variables $y_{3}\left( t\right) $ and $y_{4}\left( t\right) $ evolve
according to (\ref{A2SolvSyst}), the \textit{explicit} solution of which is
given by the relevant formulas in \textbf{Subsection} \textbf{Case A.2} of
\textbf{Appendix A} with $\tilde{m}_{1}=3,$ $\tilde{m}_{2}=4,$ $L=3,$ $%
\alpha _{0}=0~,$ $\alpha _{1}=3a,$ $\alpha _{2}=-3b,$ $\alpha _{3}=3c,$ $%
\beta _{1}=4a,$ $\beta _{2}=-4b,~\beta _{3}=4c,~\gamma _{\ell }=0.$

\bigskip

\subsection{Model 4.(ii)3.4c}

\begin{equation}
\dot{x}_{n}=x_{n}\left( a+bX+cX^{2}\right) ~,~~~X\equiv x_{1}x_{2}\left(
x_{1}+x_{2}\right) ~~~n=1,2~;  \label{4ii34cc}
\end{equation}%
$x_{1}\left( t\right) $ and $x_{2}\left( t\right) $ are related to $%
y_{3}\left( t\right) $ and $y_{4}\left( t\right) $ by (\ref{M4N2mu22ym});
and the variables $y_{3}\left( t\right) $ and $y_{4}\left( t\right) $ evolve
according to (\ref{A2SolvSyst}), the \textit{explicit} solution of which is
given by the relevant formulas in \textbf{Subsection} \textbf{Case A.2} of
\textbf{Appendix A} with $\tilde{m}_{1}=3,$ $\tilde{m}_{2}=4,$ $L=3,$ $%
\alpha _{0}=0~,$ $\alpha _{1}=3a,$ $\alpha _{2}=-\left( 3/2\right) b,$ $%
\alpha _{3}=\left( 3/4\right) c,$ $\beta _{1}=4a,$ $\beta _{2}=-2b,$ $\beta
_{3}=c,$ $\gamma _{\ell }=0.$

\bigskip

\section{Extensions}

In this \textbf{Section 5} we tersely indicate the possibility to generalize
the class of \textit{solvable} models listed in the preceding \textbf{%
Section 4}, by outlining the procedure to do so in just one case, that
detailed in \textbf{Subsection 4.3 }(see (\ref{4i12bb})), in fact just the
special case of it with $a_{0}=a_{1}=a_{3}=b_{1}=b_{3}=0$, so that its
equations of motion read as follows:
\begin{subequations}
\label{5xndotRev}
\begin{equation}
\dot{x}_{n}=\left( a_{2}x_{n}+b_{2}X\right) ~X,~~~X\equiv
3x_{1}+x_{2},~~~n=1,2~,  \label{5xndotaa}
\end{equation}%
namely
\begin{equation}
\dot{x}_{n}=c_{n1}\left( x_{1}\right) ^{2}+c_{n2}\left( x_{2}\right)
^{2}+c_{n3}x_{1}x_{2}~,~~~n=1,2~,  \label{5xndota}
\end{equation}%
with the $6$ parameters $c_{nm},$ $n=1,2,$ $m=1,2,3$, expressed as follows
in terms of the $2$ \textit{a priori arbitrary }parameters $a_{2}$ and $%
b_{2} $ (see (\ref{4i12bb})):%
\begin{eqnarray}
c_{11} &=&3\left( a_{2}+3b_{2}\right)
~,~~~c_{12}=b_{2}~,~~~c_{13}=a_{2}+6b_{2}~,  \notag \\
c_{21} &=&9b_{2}~,~~~c_{22}=a_{2}+b_{2}~,~~~c_{23}=3\left(
a_{2}+2b_{2}\right) ~.  \label{5xndotb}
\end{eqnarray}%
The \textit{explicit} solution of the initial-values problem of this system (%
\ref{5xndotRev}) is provided by \textbf{Remark A.2-1} (see \textbf{%
Subsection A.2} of \textbf{Appendix A}).

\textbf{Remark 5-1}. Note that the right-hand sides of the $2$ ODEs (\ref%
{5xndotRev}) are \textit{homogeneous polynomials of second degree}, the
coefficients of which satisfy of course the condition
\end{subequations}
\begin{equation}
\sum_{\ell =1}^{3}\left( c_{1\ell }\right) =\sum_{\ell =1}^{3}\left(
c_{2\ell }\right) ~,  \label{5CC}
\end{equation}%
as implied by \textbf{Remark 4-1}. Moreover---as clearly implied by (\ref%
{5xndotaa})---the $2$ homogeneous second-degree polynomials in the
right-hand sides of the 2 ODEs characterizing this model feature a \textit{%
common} zero: they \textit{both} vanish when $X=0$, namely when $%
x_{2}=-3x_{1}$. $\blacksquare $

Analogous extensions of other models treated in this paper shall be
performed by practitioners interested in these systems of ODEs in the
context of specific applications (see \textbf{Section 6}).

\textbf{Remark 5-2}. Note that, via a by now well-known trick (see, for
instance, \cite{C2018}) corresponding to the following time-dependent change
of both independent and dependent variables,
\begin{equation}
\tau =\exp \left( at\right) ~;~~~X_{n}\left( t\right) =\exp \left( at\right)
x_{n}\left( \tau \right) ~,~~~n=1,2~,  \label{5Xx}
\end{equation}%
the \textit{autonomous} system (\ref{5xndotRev}) gets replaced by the
following, also \textit{autonomous}, system:%
\begin{equation}
\dot{X}_{n}=aX_{n}+c_{n1}\left( X_{1}\right) ^{2}+c_{n2}\left( X_{2}\right)
^{2}+c_{n3}X_{1}X_{2}~,~~~n=1,2~.  \label{5Isochr}
\end{equation}%
Here $a$ is an \textit{arbitrary} (time-independent) parameter; and note
that if this parameter $a$ is purely \textit{imaginary}, $\func{Re}\left[ a%
\right] =0,$ $\func{Im}\left[ a\right] \neq 0,$ then this dynamical system (%
\ref{5Isochr}) is generally \textit{doubly periodic}\textbf{;} or even---if $%
a_{2}/b_{2}$ is a \textit{real rational} number---\textit{isochronous},
namely then \textit{all} its solutions are \textit{completely periodic} with
a period (an integer multiple of $T=2\pi /\left\vert a\right\vert $)
independent of the initial data: see \textbf{Remark A.2-1} and, if need be,
\cite{C2018} \cite{GS2005}. $\blacksquare $

Because of this remarkable fact, in the remaining part of this \textbf{%
Section 5} we limit, for simplicity, consideration to the special case (\ref%
{5xndotRev}), by investigating its extension which obtains via the following
linear reshuffle of the $2$ dependent variables $x_{1}\left( t\right) $ and $%
x_{2}\left( t\right) $:
\begin{subequations}
\label{5zx}
\begin{equation}
z_{1}=A_{11}x_{1}+A_{12}x_{2}~,~~~z_{2}=A_{21}x_{1}+A_{22}x_{2}~,
\label{5zxa}
\end{equation}%
which is inverted to read as follows%
\begin{equation}
x_{1}=\left( A_{22}z_{1}-A_{12}z_{2}\right) /D~,~~~x_{2}=\left(
-A_{21}z_{1}+A_{11}z_{2}\right) /D~;  \label{5xz}
\end{equation}%
here and hereafter
\begin{equation}
D=A_{11}A_{22}-A_{12}A_{21}~.  \label{5D}
\end{equation}%
It is easily seen that the new system then reads
\end{subequations}
\begin{equation}
\dot{z}_{n}=a_{n1}\left( z_{1}\right) ^{2}+a_{n2}\left( z_{2}\right)
^{2}+a_{n3}z_{1}z_{2}~,~~~n=1,2~,  \label{5zndot}
\end{equation}%
with the $6$ parameters $a_{n\ell }$, $n=1,2,$ $\ell =1,2,3$ explicitly
expressed in terms of the $4$ \textit{arbitrary} parameters $A_{nm},$\ $%
n=1,2,$ $m=1,2,$ and the $2$ \textit{arbitrary} parameters $a_{2}$ and $%
b_{2} $ (see (\ref{5xndotb})) as follows:
\begin{subequations}
\label{5an}
\begin{eqnarray}
a_{n1} &=&D^{-2}\left[ \left( A_{22}\right) ^{2}\left(
A_{n1}c_{11}+A_{n2}c_{21}\right) +\left( A_{21}\right) ^{2}\left(
A_{n1}c_{12}+A_{n2}c_{22}\right) \right.  \notag \\
&&\left. -A_{22}A_{21}\left( A_{n1}c_{13}+A_{n2}c_{23}\right) \right]
~,~~~n=1,2~,  \label{5an1}
\end{eqnarray}%
\begin{eqnarray}
a_{n2} &=&D^{-2}\left[ \left( A_{12}\right) ^{2}\left(
A_{n1}c_{11}+A_{n2}c_{21}\right) +\left( A_{11}\right) ^{2}\left(
A_{n1}c_{12}+A_{n2}c_{22}\right) \right.  \notag \\
&&\left. -A_{11}A_{12}\left( A_{n1}c_{13}+A_{n2}c_{23}\right) \right]
~,~~~n=1,2~,
\end{eqnarray}%
\begin{eqnarray}
a_{n3} &=&D^{-2}\left[ -2A_{12}A_{22}\left( A_{n1}c_{11}+A_{n2}c_{21}\right)
-2A_{21}A_{11}\left( A_{n1}c_{12}+A_{n2}c_{22}\right) \right.  \notag \\
&&\left. +\left( A_{11}A_{22}+A_{12}A_{21}\right) \left(
A_{n1}c_{13}+A_{n2}c_{23}\right) \right] ~,~~~n=1,2~.
\end{eqnarray}

\textbf{Remark 5.3}. The fact that the $6$ parameters $a_{n\ell }$ which
characterize the system (\ref{5zndot}) can be (\textit{explicitly}!)
expressed, see (\ref{5an}), in terms of $6$ \textit{a priori arbitrary}
parameters---the $4$ parameters $A_{nm},$ see (\ref{5zx}), and the $2$
parameters $a_{2}$ and $b_{2}$ (see (\ref{5xndotRev}))---might seem to imply
that this system (\ref{5zndot}) can be reduced by \textit{algebraic}
operations to the \textit{algebraically solvable} system (\ref{5xndotRev}%
)--- hence that it is itself \textit{algebraically solvable---}for \textit{%
any generic} assignment of its $6$ parameters $a_{n\ell }$, $n=1,2,$ $\ell
=1,2,3$. That this is \textit{not} the case is however implied by the
observation that the property of the system (\ref{5xndotRev})---to feature
in the right-hand sides of its $2$ ODEs $2$ polynomials themselves featuring
a \textit{common} zero (see \textbf{Remark 5-1})---is then clearly also
featured by the generalized system (\ref{5zndot}) (we like to thank Fran\c{c}%
ois Leyvraz for this very useful observation). Hence only (at most) $5$ of
the $6$ parameters $a_{n\ell }$ ($n=1,2;$ $\ell =1,2,3$) can be \textit{%
arbitrarily} assigned, since these $6$ parameters are constrained by the
condition
\end{subequations}
\begin{equation}
\left( a_{11}a_{22}-a_{21}a_{12}\right) ^{2}+\left(
a_{13}a_{21}-a_{11}a_{23}\right) \left( a_{13}a_{22}-a_{12}a_{23}\right) =0
\label{Conda}
\end{equation}%
which is easily seen to correspond to the requirement that the right-hand
sides of the $2$ ODEs (\ref{5zndot}) (with $n=1,2$) feature a \textit{common}
zero. $\blacksquare $

\textbf{Remark 5-4}. Let us finally emphasize that the trick reported in
\textbf{Remark 5-1} is just as applicable to the more general system (\ref%
{5zndot}), implying---via the \textit{ansatz}
\begin{subequations}
\begin{equation}
\tau =\exp \left( at\right) ~;~~~Z_{n}\left( t\right) =\exp \left( at\right)
z_{n}\left( \tau \right) ~,~~~n=1,2~,  \label{5Zz}
\end{equation}%
analogous to (\ref{5Xx})---the solvability of the system%
\begin{equation}
\dot{Z}_{n}=aZ_{n}+a_{n1}\left( Z_{1}\right) ^{2}+a_{n2}\left( Z_{2}\right)
^{2}+a_{n3}Z_{1}Z_{2}~,~~~n=1,2~,  \label{52zndot}
\end{equation}%
featuring the $7$ parameters $a$ and $a_{n\ell }$ ($n=1,2;$ $\ell =1,2,3$). $%
\blacksquare $

The relevance of this dynamical system, (\ref{52zndot}), in many applicative
context is exemplified by too many contributions to allow reporting a full
bibliography; we record here just one such paper which lists $11$ references
and contains the remarkable assertion that the system (\ref{52zndot}) "is
not solvable explicitly except in certain simple cases" \cite{N2005}.

\bigskip

\section{Outlook}

In this final \textbf{Section 6} we tersely outline future developments of
the findings reported in this paper.

There is of course the possibility to treat cases with $M>4$ (see \textbf{%
Section 3}).

There is the possibility to \textit{iterate} the procedure leading to the
identification of \textit{new solvable} systems (as described in this
paper): see for this kind of development \cite{BC2016} and Chapter 6 of \cite%
{C2018}.

Another natural development is to treat analogous dynamical systems evolving
in \textit{discrete }rather than \textit{continuous} time. For progress in
this direction see \cite{CP2019}.

Another extension is to treat systems characterized by \textit{second-order}
rather than \textit{first-order} differential equations, including models
characterized by Newtonian equations of motion ("accelerations equal
forces"); and in the cases in which these equations of motion are derivable
from a Hamiltonian, an additional interesting development is the treatment
of the corresponding time-evolutions in the context of \textit{quantal}
rather than \textit{classical} mechanics.

And yet another extension is to Partial Differential Equations (PDEs) rather
than ODEs.

There is finally the vast universe of applications, including to cases in
which the systems of evolution equations can be shown---via their \textit{%
solvability}---to feature remarkable properties such as \textit{isochrony}
\cite{C2008} \cite{GS2005} or \textit{asymptotic isochrony} \cite{CG2008}.

\bigskip

\section{Acknowledgements}

FP likes to thank the Physics Department of the University of Rome\ "La
Sapienza" for the hospitality from April to November 2018 (during her
sabbatical), when the results reported in this paper were obtained. FC likes
to thank Robert Conte and Fran\c{c}ois Leyvraz for very useful discussions
in the context of the Gathering of Scientists on "Integrable systems and
beyond" hosted by the Centro Internacional de Ciencias (CIC) in Cuernavaca,
Mexico, from November 19th to December 14th, 2018.

\bigskip

\section{Appendix A: Three useful classes of solvable systems of $2$
nonlinear first-order ODEs for the $2$ variables $y_{\tilde{m}}\left(
t\right) $}

The findings reported in this \textbf{Appendix A} are not new; they are
displayed here to facilitate the reader of the new findings reported in the
body of this paper.

\textbf{Notation A-1}. In this \textbf{Appendix A} we indicate with the
notation $y_{\tilde{m}_{1}}\left( t\right) $ and $y_{\tilde{m}_{2}}\left(
t\right) $---with $\tilde{m}_{1,2}=1,2,3,4$ and $\tilde{m}_{1}\neq \tilde{m}%
_{2}$ (and for the significance of the superimposed tilde see the last part
of \textbf{Section 2})---the $2$ dependent variables which satisfy the
"solvable" system of $2$ nonlinearly-coupled ODEs
\end{subequations}
\begin{equation}
\dot{y}_{\tilde{m}_{1}}=f_{\tilde{m}_{1}}\left( y_{\tilde{m}_{1}},y_{\tilde{m%
}_{2}}\right) ~,~~~\dot{y}_{\tilde{m}_{2}}=f_{\tilde{m}_{2}}\left( y_{\tilde{%
m}_{1}},y_{\tilde{m}_{2}}\right) ~,  \label{ASolvSyst}
\end{equation}%
with the $2$ functions $f_{\tilde{m}_{1}}\left( y_{\tilde{m}_{1}},y_{\tilde{m%
}_{2}}\right) $ and $f_{\tilde{m}_{2}}\left( y_{\tilde{m}_{1}},y_{\tilde{m}%
_{2}}\right) $ assigned---conveniently for our treatment in this paper (see
\textbf{Section 2} above)---so that the system (\ref{ASolvSyst}) is
"solvable". The precise meaning of the term "solvable" shall be clear from
the following.

In this \textbf{Appendix} \textbf{A} $\alpha _{\ell },$ $\beta _{\ell },$ $%
\gamma _{\ell }$ are \textit{a priori arbitrary} \textit{time-independent}
parameters, and $L$ is an \textit{a priori arbitrary nonnegative integer}.

The selection of the specific systems of $2$ ODEs considered below is of
course motivated by the treatment in the body of this paper, see in
particular \textbf{Sections 2} and \textbf{3}. $\blacksquare $

\bigskip

\subsection{Case A.1}

\begin{equation}
\dot{y}_{\tilde{m}_{1}}=\sum_{\ell =0}^{L}\left[ \alpha _{\ell }\left( y_{%
\tilde{m}_{1}}\right) ^{\ell \tilde{m}_{2}+1}\right] ~,~~~\dot{y}_{\tilde{m}%
_{2}}=\sum_{\ell =0}^{L}\left[ \beta _{\ell }\left( y_{\tilde{m}_{2}}\right)
^{\ell \tilde{m}_{1}+1}\right] ~.  \label{A1SolvSyst}
\end{equation}%
Each of these $2$ ODEs can be integrated via one \textit{quadrature}, that
can be performed explicitly after some purely \textit{algebraic} operations.
Indeed, to integrate the \textit{first} of these $2$ ODEs one must first of
all identify---via an \textit{algebraic} operation---the $L\tilde{m}_{2}+1$
zeros $\bar{y}_{n}$ (assumed below, for simplicity, to be all different
among themselves) of the polynomial in its right-hand side,%
\begin{equation}
\sum_{\ell =0}^{L}\left[ \alpha _{\ell }\left( y_{\tilde{m}_{1}}\right)
^{\ell \tilde{m}_{2}+1}\right] =\alpha _{L}\prod\limits_{n=1}^{L\tilde{m}%
_{2}+1}\left( y_{\tilde{m}_{1}}-\bar{y}_{n}\right) ~;
\end{equation}%
next one must identify the $L\tilde{m}_{2}+1$ "residues" $r_{n}$ defined by
the "partial fraction decomposition" formula%
\begin{equation}
\prod\limits_{n=1}^{L\tilde{m}_{2}+1}\left( y_{\tilde{m}_{1}}-\bar{y}%
_{n}\right) ^{-1}=\sum_{n=1}^{L\tilde{m}_{2}+1}\left[ r_{n}\left( y_{\tilde{m%
}_{1}}-\bar{y}_{n}\right) ^{-1}\right]
\end{equation}%
---another \textit{algebraic} operation, indeed one that can be performed
\textit{explicitly}; and finally one integrates the ODE getting the
(generally \textit{implicit;} but not always, see below) result%
\begin{equation}
\prod\limits_{n=1}^{L\tilde{m}_{2}+1}\left\{ \left[ \frac{y_{\tilde{m}%
_{1}}\left( t\right) -\bar{y}_{n}}{y_{\tilde{m}_{1}}\left( 0\right) -\bar{y}%
_{n}}\right] ^{r_{n}}\right\} =\exp \left( \frac{t}{\alpha _{L}}\right) ~,
\end{equation}%
which characterizes the solution $y_{\tilde{m}_{1}}\left( t\right) $
corresponding to the initial datum $y_{\tilde{m}_{1}}\left( 0\right) .$

Of course an analogous procedure characterizes---for the second ODE (\ref%
{A1SolvSyst})---the solution $y_{\tilde{m}_{2}}\left( t\right) $
corresponding to the initial datum $y_{\tilde{m}_{2}}\left( 0\right) .$

For $L=1$ the initial-value problem for these $2$ ODEs can be solved \textit{%
explicitly}, since the solution of the initial-value problem for the ODE
\begin{subequations}
\begin{equation}
\dot{y}=Ay+By^{M+1}
\end{equation}%
is provided by the formula%
\begin{equation}
y\left( t\right) =y\left( 0\right) \exp \left( At\right) \left\{ 1+\left(
B/A\right) \left[ y\left( 0\right) \right] ^{M}\left[ 1-\exp \left(
MAt\right) \right] \right\} ^{-1}~.
\end{equation}

\bigskip

\subsection{Case A.2}

\end{subequations}
\begin{subequations}
\label{A2SolvSyst}
\begin{equation}
\dot{y}_{\tilde{m}_{1}}=\sum_{\ell =0}^{L}\left[ \alpha _{\ell }\left( y_{%
\tilde{m}_{1}}\right) ^{\ell }\right] ~,  \label{A2SolvSysta}
\end{equation}%
\begin{equation}
\dot{y}_{\tilde{m}_{2}}=y_{\tilde{m}_{2}}\sum_{\ell =1}^{L}\left[ \beta
_{\ell }\left( y_{\tilde{m}_{1}}\right) ^{\ell -1}\right] +\sum_{\ell =0}^{L}%
\left[ \gamma _{\ell }\left( y_{\tilde{m}_{1}}\right) ^{\ell -1+\left(
\tilde{m}_{2}/\tilde{m}_{1}\right) }\right] ~.  \label{A2SolvSystb}
\end{equation}

The solution of the first of these $2$ ODEs, (\ref{A2SolvSysta}), has been
already discussed above, see \textbf{Subsection} \textbf{Case A.1}; hence in
this \textbf{Subsection Case A.2} we need to consider only the second ODE (%
\ref{A2SolvSystb}).$\ $ And since we are mainly interested in the case when
the right-hand side of this ODE is polynomial, we shall limit our
consideration below to the $3$ subcases with $\tilde{m}_{1}=1$ and to the
single case with $\tilde{m}_{1}=2,$ $\tilde{m}_{2}=4;$ except in the special
case with all parameters $\gamma _{\ell }$ vanishing, $\gamma _{\ell }=0$,
which we treat separately \textit{firstly} (since it is an intermediate step
to solve the more general case).

In this special case the ODE (\ref{A2SolvSystb}) reads
\end{subequations}
\begin{subequations}
\begin{equation}
\dot{y}_{\tilde{m}_{2}}=y_{\tilde{m}_{2}}\sum_{\ell =1}^{L}\left[ \beta
_{\ell }\left( y_{\tilde{m}_{1}}\right) ^{\ell -1}\right] ~,
\end{equation}%
with the function $y_{\tilde{m}_{1}}\left( t\right) $ to be considered
known; hence the solution of the initial-value problem for this ODE reads%
\begin{equation}
y_{\tilde{m}_{2}}\left( t\right) =y_{\tilde{m}_{2}}\left( 0\right) ~F\left(
t\right)
\end{equation}%
with%
\begin{equation}
F\left( t\right) =\exp \left\{ \int_{0}^{t}\left[ dt^{\prime }\sum_{\ell
=1}^{L}\left\{ \beta _{\ell }\left[ y_{\tilde{m}_{1}}\left( t^{\prime
}\right) \right] ^{\ell -1}\right\} \right] \right\} ~.  \label{A2F}
\end{equation}

And it is then easily seen that the solution of the initial-value problem of
the (more general) ODE (\ref{A2SolvSystb}) reads
\end{subequations}
\begin{equation}
y_{\tilde{m}_{2}}\left( t\right) =F\left( t\right) \left[ y_{\tilde{m}%
_{2}}\left( 0\right) +\int_{0}^{t}dt^{\prime }~\left[ F\left( t^{\prime
}\right) \right] ^{-1}\sum_{\ell =0}^{L}\left\{ \gamma _{\ell }\left[ y_{%
\tilde{m}_{1}}\left( t^{\prime }\right) \right] ^{\ell -1+\left( \tilde{m}%
_{2}/\tilde{m}_{1}\right) }\right\} \right] ~.
\end{equation}

More explicit solutions can be easily obtained in the following cases:
\begin{subequations}
\begin{equation}
\tilde{m}_{1}=1~,~~~\tilde{m}_{2}=2,3,4~~~\text{or~~~}\tilde{m}_{1}=2,~~~%
\tilde{m}_{2}=4~,~~~M=\tilde{m}_{2}/\tilde{m}_{1}~,
\end{equation}%
\begin{equation}
\dot{y}_{\tilde{m}_{1}}=\alpha _{0}+\alpha _{1}y_{\tilde{m}_{1}}+\alpha
_{2}\left( y_{\tilde{m}_{1}}\right) ^{2}~,
\end{equation}%
\begin{equation}
\dot{y}_{\tilde{m}_{2}}=y_{\tilde{m}_{2}}\left( \beta _{0}+\beta _{1}y_{%
\tilde{m}_{1}}\right) +\gamma _{0}\left( y_{\tilde{m}_{1}}\right)
^{-1+M}+\gamma _{1}\left( y_{\tilde{m}_{1}}\right) ^{M}+\gamma _{2}\left( y_{%
\tilde{m}_{1}}\right) ^{1+M}~;
\end{equation}%
\begin{eqnarray}
y_{_{\tilde{m}_{1}}}\left( t\right) &=&\frac{y_{_{\tilde{m}_{1}}}\left(
0\right) \left[ 1+\left( \Delta /\alpha _{1}\right) \tanh \left( \Delta
t\right) \right] -2\left( \alpha _{0}/\alpha _{1}\right) \tanh \left( \Delta
t\right) }{1-\left\{ \left[ 2\alpha _{2}y_{_{\tilde{m}_{1}}}\left( 0\right)
+\Delta \right] /\alpha _{1}\right\} \tanh \left( \Delta t\right) }~,  \notag
\\
\Delta ^{2} &=&\left( \alpha _{1}\right) ^{2}-4\alpha _{0}\alpha _{2}~,
\end{eqnarray}%
\end{subequations}
\begin{subequations}
\begin{equation}
y_{\tilde{m}_{2}}\left( t\right) =f\left( t\right) \left[ y_{\tilde{m}%
_{2}}\left( 0\right) +\int_{0}^{t}dt^{\prime }~\left[ f\left( t^{\prime
}\right) \right] ^{-1}\sum_{\ell =0}^{L}\left\{ \gamma _{\ell }\left[ y_{%
\tilde{m}_{1}}\left( t^{\prime }\right) \right] ^{\ell -1+\left( \tilde{m}%
_{2}/\tilde{m}_{1}\right) }\right\} \right] ~,
\end{equation}%
\begin{equation}
f\left( t\right) =\exp \left\{ \int_{0}^{t}\left[ dt^{\prime }\sum_{\ell
=1}^{L}\left\{ \beta _{\ell }\left[ y_{\tilde{m}_{1}}\left( t^{\prime
}\right) \right] ^{\ell -1}\right\} \right] \right\} ~.
\end{equation}%
\

\textbf{Remark A.2-1}. In particular, it is easily seen that the system of
ODEs (\ref{5xndotRev}) discussed in \textbf{Section 5} (see \textbf{%
Subsection 4.3 }with $a_{0}=a_{1}=a_{3}=b_{1}=b_{3}=0$) implies that the
system of ODEs (\ref{A2SolvSyst}), which then reads
\end{subequations}
\begin{subequations}
\begin{equation}
\dot{y}_{1}=\alpha _{2}\left( y_{1}\right) ^{2}~,~~~\dot{y}_{2}=\beta
_{2}y_{2}y_{1}+\gamma _{2}\left( y_{1}\right) ^{3}~,
\end{equation}%
with%
\begin{equation}
\alpha _{2}=-\left( a_{2}+4b_{2}\right) ~,~~~\beta _{2}=-2a_{2}~,~~~\gamma
_{2}=3b_{2}~,
\end{equation}%
features the following \textit{explicit} solution of its initial-values
problem:
\end{subequations}
\begin{subequations}
\label{A2Remark1}
\begin{equation}
y_{1}\left( t\right) =\frac{y_{1}\left( 0\right) }{1-\alpha _{2}y_{1}\left(
0\right) t}~,
\end{equation}%
\begin{eqnarray}
y_{2}\left( t\right) &=&\left\{ y_{2}\left( 0\right) +\left( 3/8\right)
\left[ y_{1}\left( 0\right) \right] ^{2}\right\} \left[ 1-\alpha
_{2}y_{1}\left( 0\right) t\right] ^{-\beta _{2}/\alpha _{2}}  \notag \\
&&-\left( \frac{3}{8}\right) \left[ \frac{y_{1}\left( 0\right) }{1-\alpha
_{2}y_{1}\left( 0\right) t}\right] ^{2}~.
\end{eqnarray}%
The corresponding solution of the initial-values problem for the system of
ODEs (\ref{5xndotRev}) is then obtained from this solution (\ref{A2Remark1})
via the relations $y_{1}=-3x_{1}-x_{2}$,~$y_{2}=3x_{1}\left(
x_{1}+x_{2}\right) $ (see (\ref{M4N2mu31ym})), which of course imply
\end{subequations}
\begin{subequations}
\begin{equation}
x_{2}\left( t\right) =-3x_{1}\left( t\right) -y_{1}\left( t\right) ~,
\end{equation}%
with $x_{1}\left( t\right) $ given in terms of $y_{1}\left( t\right) $ and $%
y_{2}\left( t\right) $ (see (\ref{A2Remark1})) as a solution of the (\textit{%
explicitly solvable}!) second-degree equation%
\begin{equation}
6\left( x_{1}\right) ^{2}+3y_{1}x_{1}+y_{2}=0~.~~~\blacksquare
\end{equation}

\bigskip

\subsection{Case A.3}

\end{subequations}
\begin{subequations}
\begin{equation}
\dot{y}_{\tilde{m}_{1}}=\alpha _{0}+\alpha _{1}y_{\tilde{m}_{2}}~,~~~\dot{y}%
_{\tilde{m}_{2}}=\beta _{0}\left( y_{\tilde{m}_{1}}\right) ^{-1+\left(
\tilde{m}_{2}/\tilde{m}_{1}\right) }+\beta _{1}\left( y_{\tilde{m}%
_{1}}\right) ^{-1+2\left( \tilde{m}_{2}/\tilde{m}_{1}\right) }~.
\label{A3SolvSysta}
\end{equation}%
Since we are mainly interested in the case when the right-hand sides of both
these ODEs are polynomial, we shall limit our consideration of this case to
the $3$ subcases with $\tilde{m}_{1}=1$ and to the single case with $\tilde{m%
}_{1}=2,$ $\tilde{m}_{2}=4.$

The most convenient technique to solve the system of $2$ ODEs (\ref%
{A3SolvSysta}) is by noticing to begin with that it implies for the variable
$y_{\tilde{m}_{1}}\left( t\right) $ the second-order ODE%
\begin{equation}
\ddot{y}_{\tilde{m}_{1}}=\alpha _{1}\left[ \beta _{0}\left( y_{\tilde{m}%
_{1}}\right) ^{-1+\left( \tilde{m}_{2}/\tilde{m}_{1}\right) }+\beta
_{1}\left( y_{\tilde{m}_{1}}\right) ^{-1+2\left( \tilde{m}_{2}/\tilde{m}%
_{1}\right) }\right] ~,  \label{A3SolvSystb}
\end{equation}%
which is an \textit{autonomous} Newtonian equation of motion ("acceleration
equal force") and is of course solvable by \textit{quadratures} (as
discussed in more detail in the following special cases).

And of course once $y_{\tilde{m}_{1}}\left( t\right) $ is known, $\dot{y}_{%
\tilde{m}_{2}}=\left[ \dot{y}_{\tilde{m}_{1}}\left( t\right) -\alpha _{0}%
\right] /\alpha _{1}$ is as well known (see the first of the $2$ ODEs (\ref%
{A3SolvSysta})).

\bigskip

\subsubsection{Case A.3.1}

In this case $\tilde{m}_{1}=1,$ $\tilde{m}_{2}=2,$ or $\tilde{m}_{1}=2,$ $%
\tilde{m}_{2}=4,$ so that (\ref{A3SolvSystb}) reads
\end{subequations}
\begin{equation}
\ddot{w}=\alpha _{1}\left( \beta _{0}w+\beta _{1}w^{3}\right) ~;
\end{equation}%
here and below $w\left( t\right) $ stands for $y_{\tilde{m}_{1}}\left(
t\right) $ respectively $y_{\tilde{m}_{2}}\left( t\right) ,$ in the $2$
cases $\tilde{m}_{1}=1,$ $\tilde{m}_{2}=2,$ respectively $\tilde{m}_{1}=2,$ $%
\tilde{m}_{2}=4$.

It is easily seen that the solution of the initial-value problem of this ODE
reads as follows:
\begin{subequations}
\begin{equation}
w\left( t\right) =\mu ~\text{sn}\left( \lambda ~t+\rho ,k\right) ~,
\label{A1a1w1}
\end{equation}%
with the $4$ parameters $\lambda ,$ $\mu ,$ $\rho $ and $k$ determined by
the following $4$ formulas in terms of the $3$ parameters $\alpha _{1},$ $%
\beta _{0},$ $\beta _{1}$ and the initial data $w\left( 0\right)
=y_{1}\left( 0\right) $, $u\left( 0\right) =y_{2}\left( 0\right) $
respectively $w\left( 0\right) =y_{2}\left( 0\right) $, $u\left( 0\right)
=y_{4}\left( 0\right) $ (in the $2$ cases $\tilde{m}_{1}=1,$ $\tilde{m}%
_{2}=2,$ respectively $\tilde{m}_{1}=2,$ $\tilde{m}_{2}=4$):%
\begin{eqnarray}
\lambda ^{2}=\frac{-\alpha _{1}\beta _{0}\beta _{1}}{1+k^{2}}~,~~~\mu ^{2}=%
\frac{-2k^{2}\beta _{0}}{\beta _{1}\left( 1+k^{2}\right) }~,~~~\text{sn}%
\left( \rho ,k\right) =\frac{w\left( 0\right) }{\mu }~, &&  \notag \\
c_{1}k^{2}+c_{2}\left( 1+k^{2}\right) ^{2}=0~,~~~c_{1}=-2\alpha _{1}\left(
\beta _{0}\right) ^{2}~, &&  \notag \\
c_{2}=\left( \alpha _{1}\right) ^{2}\left[ \beta _{0}+\beta _{1}u\left(
0\right) \right] ^{2}-\alpha _{1}\beta _{0}\beta _{1}\left[ w\left( 0\right) %
\right] ^{2}-\frac{\alpha _{1}\left( \beta _{1}\right) ^{2}}{2}\left[
w\left( 0\right) \right] ^{4}~. &&
\end{eqnarray}%
And of course correspondingly (see the first of the $2$ ODEs (\ref%
{A3SolvSysta}))
\end{subequations}
\begin{equation}
u\left( t\right) =\left( \alpha _{1}\beta _{1}\right) ^{-1}\left[ \lambda
\mu ~\text{cn}\left( \lambda t+\rho ,k\right) ~\text{dn}\left( \lambda
t+\rho ,k\right) -\alpha _{1}\beta _{0}\right] ~,  \label{A1a1w2}
\end{equation}%
where of course $u\left( t\right) $ stands for $y_{2}\left( t\right) $
respectively $y_{4}\left( t\right) .$

Here sn$\left( z,k\right) ,$ cn$\left( z,k\right) ,$ dn$\left( z,k\right) $
are the $3$ standard Jacobian elliptic functions.

\bigskip

\subsubsection{Case A.3.2}

In this case $\tilde{m}_{1}=1,$ $\tilde{m}_{2}=3,$ so that (\ref{A3SolvSystb}%
) reads
\begin{subequations}
\begin{equation}
\ddot{y}_{1}=\alpha _{1}\left[ \beta _{0}\left( y_{1}\right) ^{2}+\beta
_{1}\left( y_{1}\right) ^{5}\right] ~,
\end{equation}%
implying%
\begin{eqnarray}
&&\int\limits_{y_{1}\left( 0\right) }^{y_{1}\left( t\right) }\frac{dy}{\sqrt{%
C+\left( \alpha _{1}/3\right) y^{3}\left( 2\beta _{0}+\beta _{1}y^{3}\right)
}}=t~,  \notag \\
C &=&\left[ y_{1}\left( 0\right) \right] ^{2}-\left( \alpha _{1}/3\right) %
\left[ y_{1}\left( 0\right) \right] ^{3}\left\{ 2\beta _{0}+\beta _{1}\left[
y_{1}\left( 0\right) \right] ^{3}\right\} ~.
\end{eqnarray}

It is thus seen that in this case $y_{1}\left( t\right) $ is a hyperelliptic
function.

\bigskip

\subsubsection{Case A.3.3}

In this case $\tilde{m}_{1}=1,$ $\tilde{m}_{2}=4,$ so that (\ref{A3SolvSystb}%
) reads
\end{subequations}
\begin{subequations}
\begin{equation}
\ddot{y}_{1}=\alpha _{1}\left[ \beta _{0}\left( y_{1}\right) ^{3}+\beta
_{1}\left( y_{1}\right) ^{7}\right] ~,
\end{equation}%
implying%
\begin{eqnarray}
&&\int\limits_{y_{1}\left( 0\right) }^{y_{1}\left( t\right) }\frac{dy}{\sqrt{%
C+\left( \alpha _{1}/4\right) y^{4}\left( 2\beta _{0}+\beta _{1}y^{4}\right)
}}=t~,  \notag \\
C &=&\left[ y_{1}\left( 0\right) \right] ^{2}-\left( \alpha _{1}/4\right) %
\left[ y_{1}\left( 0\right) \right] ^{4}\left\{ 2\beta _{0}+\beta _{1}\left[
y_{1}\left( 0\right) \right] ^{4}\right\} ~.
\end{eqnarray}

It is thus again seen that in this case $y_{1}\left( t\right) $ is a
hyperelliptic function.

\bigskip

\section{Appendix B}

In this \textbf{Appendix B} we display---for the case $M=4$ and $N=2$---the
expressions of the time-derivatives $\dot{y}_{\tilde{m}}\left( t\right) $ of
the coefficients $y_{\tilde{m}}\left( t\right) $ in terms of the
time-derivatives $\dot{y}_{m}\left( t\right) $ of the \textit{coefficients} $%
y_{m}\left( t\right) $ and of the \textit{zeros} $x_{n}\left( t\right) $,
for the $2$ cases with $\mu _{1}=3,$ $\mu _{2}=1$ respectively $\mu _{1}=\mu
_{2}=2$ (for this terminology, see \textbf{Section 2}).

In case (i), with $\mu _{1}=3,$ $\mu _{2}=1,$ these relations read as
follows:
\end{subequations}
\begin{subequations}
\begin{equation}
\dot{y}_{1}=-\frac{2x_{1}\dot{y}_{2}+\dot{y}_{3}}{3\left( x_{1}\right) ^{3}}%
~,~~~\dot{y}_{1}=-\frac{\left( x_{1}\right) ^{2}\dot{y}_{2}-\dot{y}_{4}}{%
2\left( x_{1}\right) ^{3}}~,~~~\dot{y}_{1}=\frac{x_{1}\dot{y}_{3}+2\dot{y}%
_{4}}{\left( x_{1}\right) ^{3}}~,
\end{equation}%
\begin{equation}
\dot{y}_{2}=-\frac{3\left( x_{1}\right) ^{2}\dot{y}_{1}+\dot{y}_{3}}{2x_{1}}%
~,~~~\dot{y}_{2}=-\frac{2\left( x_{1}\right) ^{3}\dot{y}_{1}-\dot{y}_{4}}{%
\left( x_{1}\right) ^{2}}~,~~~\dot{y}_{2}=-\frac{2x_{1}\dot{y}_{3}+3\dot{y}%
_{4}}{\left( x_{1}\right) ^{2}}~,
\end{equation}%
\begin{equation}
\dot{y}_{3}=-3\left( x_{1}\right) ^{2}\dot{y}_{1}-2x_{1}\dot{y}_{2}~,~~~\dot{%
y}_{3}=\frac{\left( x_{1}\right) ^{3}\dot{y}_{1}-2\dot{y}_{4}}{x_{1}}~,~~~%
\dot{y}_{3}=-\frac{\left( x_{1}\right) ^{2}\dot{y}_{2}+3\dot{y}_{4}}{2x_{1}}%
~,
\end{equation}%
\begin{eqnarray}
\dot{y}_{4} &=&2\left( x_{1}\right) ^{3}\dot{y}_{1}+\left( x_{1}\right) ^{2}%
\dot{y}_{2}~,~~\hspace{0in}\dot{y}_{4}=\frac{1}{2}\left[ \left( x_{1}\right)
^{3}\dot{y}_{1}-x_{1}\dot{y}_{3}\right] ~,  \notag \\
\dot{y}_{4} &=&-\frac{1}{3}\left[ \left( x_{1}\right) ^{2}\dot{y}_{2}+2x_{1}%
\dot{y}_{3}\right] ~.
\end{eqnarray}

In case (ii), with $\mu _{1}=\mu _{2}=2,$ these relations read as follows:
\end{subequations}
\begin{subequations}
\begin{eqnarray}
\dot{y}_{1} &=&-\frac{\left( x_{1}+x_{2}\right) \dot{y}_{2}+\dot{y}_{3}}{%
\left( x_{1}\right) ^{2}+x_{1}x_{2}+\left( x_{2}\right) ^{2}}~,~~~\dot{y}%
_{1}=-\frac{x_{1}x_{2}\dot{y}_{2}-\dot{y}_{4}}{x_{1}x_{2}\left(
x_{1}+x_{2}\right) }~,  \notag \\
\dot{y}_{1} &=&\frac{x_{1}x_{2}\dot{y}_{3}+\left( x_{1}+x_{2}\right) \dot{y}%
_{4}}{\left( x_{1}x_{2}\right) ^{2}}~,
\end{eqnarray}%
\begin{eqnarray}
\dot{y}_{2} &=&-\frac{\left[ \left( x_{1}\right) ^{2}+x_{1}x_{2}+\left(
x_{2}\right) ^{2}\right] \dot{y}_{1}+\dot{y}_{3}}{x_{1}+x_{2}}~,~~~\dot{y}%
_{2}=-\frac{x_{1}x_{2}\left( x_{1}+x_{2}\right) \dot{y}_{1}-\dot{y}_{4}}{%
x_{1}x_{2}}~,  \notag \\
\dot{y}_{2} &=&-\frac{x_{1}x_{2}\left( x_{1}+x_{2}\right) \dot{y}_{3}+\left[
\left( x_{1}\right) ^{2}+x_{1}x_{2}+\left( x_{2}\right) ^{2}\right] \dot{y}%
_{4}}{\left( x_{1}x_{2}\right) ^{2}}~,
\end{eqnarray}%
\begin{eqnarray}
\dot{y}_{3} &=&-\left[ \left( x_{1}\right) ^{2}+x_{1}x_{2}+\left(
x_{2}\right) ^{2}\right] \dot{y}_{1}-\left( x_{1}+x_{2}\right) \dot{y}_{2}~,
\notag \\
\dot{y}_{3} &=&\frac{\left( x_{1}x_{2}\right) ^{2}\dot{y}_{1}-\left(
x_{1}+x_{2}\right) \dot{y}_{4}}{x_{1}x_{2}}~,  \notag \\
\dot{y}_{3} &=&-\frac{\left( x_{1}x_{2}\right) ^{2}\dot{y}_{2}+\left[ \left(
x_{1}\right) ^{2}+x_{1}x_{2}+\left( x_{2}\right) ^{2}\right] \dot{y}_{4}}{%
x_{1}x_{2}\left( x_{1}+x_{2}\right) }~,
\end{eqnarray}%
\begin{eqnarray}
\dot{y}_{4} &=&x_{1}x_{2}\left( x_{1}+x_{2}\right) \dot{y}_{1}+x_{1}x_{2}%
\dot{y}_{2}~,~~~\dot{y}_{4}=\frac{\left( x_{1}x_{2}\right) ^{2}\dot{y}%
_{1}-x_{1}x_{2}\dot{y}_{3}}{x_{1}+x_{2}}~,  \notag \\
\dot{y}_{4} &=&-\frac{\left( x_{1}x_{2}\right) ^{2}\dot{y}%
_{2}+x_{1}x_{2}\left( x_{1}+x_{2}\right) \dot{y}_{3}}{\left( x_{1}\right)
^{2}+x_{1}x_{2}+\left( x_{2}\right) ^{2}}~.
\end{eqnarray}

\end{subequations}

\end{document}